\documentclass[12pt]{article}
\usepackage{amsmath}

\newtheorem{theorem}{Theorem}

\begin{document}
 
\title{Reciprocal transformations of Hamiltonian operators of hydrodynamic
type: nonlocal Hamiltonian formalism for linearly degenerate systems}
\author{E.V.Ferapontov \& M.V.Pavlov \\
Department of Mathematical Sciences \\
Loughborough University \\
Loughborough, Leicestershire LE11 3TU \\
United Kingdom \\
e-mail: \texttt{E.V.Ferapontov@lboro.ac.uk}}
\date{}
\maketitle
 
\begin{abstract}
Reciprocal transformations of Hamiltonian operators of hydrodynamic type are
investigated. The transformed operators are generally nonlocal, possessing a
number of remarkable algebraic and differential-geometric properties. We
apply our results to linearly degenerate semi-Hamiltonian systems in Riemann
invariants, a typical example being%
\begin{equation*}
R_{t}^{i}=\left( \sum_{m=1}^n R^{m}-R^{i}\right) R_{x}^{i}\text{, \ \ }i=1,2,..., n.
\end{equation*}%
Since all such systems are linearizable by appropriate (generalized) reciprocal
transformations, our formulae provide an infinity of mutually compatible
nonlocal Hamiltonian structures, explicitly parametrized by $n$ arbitrary
functions of one variable.
 
 
\end{abstract}

\newpage
 
\section{Introduction}

Equations of hydrodynamic type,
$$
u^i_t=\upsilon^i_j(u) u^j_x, ~~~ i, j =1, ..., n,
$$
naturally arise in apllications such as gas dynamics, hydrodynamics, chemical kinetics, the Whitham averaging procedure, differential geometry and topological field theory,
see, e.g.,  \cite{dn83}, \cite{dn89}, \cite{ts85}, \cite{ts91}, \cite{d96}, \cite%
{nt}, \cite{ol}.  In this paper we study Hamiltonian systems,
\begin{equation}
u_{t}^{i}=\upsilon _{j}^{i}(u)u_{x}^{j}=J^{ij}\frac{\partial h}{\partial
u^{j}}\text{, \ \ \ }i=1,2,..., n,  \label{hydro}
\end{equation}%
where%
\begin{equation}
J^{ij}=g^{ij}(u)\frac{d}{dx}-g^{is}(u)\Gamma _{sk}^{j}(u)u_{x}^{k}
\label{ham}
\end{equation}
is the Hamiltonian operator and $h(u)$ is the density. As pointed out by
Dubrovin and Novikov \cite{dn83}, expression (\ref{ham}) defines a
Hamiltonian operator if and only if the metric $g^{ij}$ ($g^{ij}=g^{ji}$, $
\det g^{ij}\neq 0$) is flat and $\Gamma _{sk}^{j}$ are the Christoffel
symbols of the corresponding Levi-Civita connection. Notice that equations (%
\ref{hydro}) and (\ref{ham}) imply that $\upsilon _{j}^{i}$ is the matrix of
second covariant derivatives of the density $h$,%
\begin{equation*}
\upsilon _{j}^{i}=\nabla ^{i}\nabla _{j}h,
\end{equation*}%
where $\nabla ^{i}\equiv g^{ij}\nabla _{j}$.  The theory of integrability of Hamiltonian systems of hydrodynamic type was proposed  by Tsarev \cite{ts85, ts91}. He demonstrated that if a Hamiltonian system (\ref{hydro}) is diagonalizable, that is, can be written in the diagonal form
$$
R^i_t=v^i(R) R^i_x,
$$
(the corresponding coordinates $R^i$ are called the Riemann invariants), then it necessarily possesses an infinity of conservation laws 
and commuting flows and, moreover, is integrable by the so-called `generalized hodograph transform'. In Riemann invariants $R^i$, the flat metric
$g^{ij}$ also becomes diagonal, $g^{ij}=g^{ii}\delta^{ij}$, thus establishing a correspondence between Hamiltonian diagonalizable systems (\ref{hydro})  and
n-orthogonal curvilinear coordinates in flat spaces. This shows that the theory of Hamiltonian systems of hydrodynamic type is deeply rooted in classical differential geometry.

There exists an important class of non-local transformations  which act on hydrodynamic-type systems. Unlike the  classical B\"acklund transformations (Miura-type transformations) familiar from the
soliton theory, these transformations change the {\it independent} variables of a system. They are defined as follows.
Let $B(u)dx+A(u)dt$ and $N(u)dx+M(u)dt$ be two conservation laws of system (%
\ref{hydro}), understood as  one-forms which are closed by virtue of (\ref%
{hydro}). In the new independent variables $\tilde{x}$ and $\tilde{t}$
defined by
\begin{equation}
d\tilde{x}=B(u)dx+A(u)dt\text{, \ \ \ \ }d\tilde{t}=N(u)dx+M(u)dt,
\label{recip}
\end{equation}%
equations (\ref{hydro}) take the form%
\begin{equation}
u_{\tilde{t}}^{i}=\tilde{\upsilon}_{j}^{i}(u)u_{\tilde{x}}^{j},
\label{modif}
\end{equation}%
where the transformed matrix $\tilde{\upsilon}$ is $\tilde{\upsilon}%
=(B\upsilon -AE)(ME-N\upsilon )^{-1}$, $E=id$. Transformations of the type (%
\ref{recip}) originate from gas dynamics and are known as {\it reciprocal}  \cite%
{rg}. The simplest example is, probably, the passage from Eulerian to  Lagrangian coordinates in   1-dimensional gas dynamics. Reciprocal transformations are known to preserve the class of diagonalizable systems, which is an immediate corollary of the form of the transformed matrix $\tilde v$. Moreover, all conservation laws and commuting flows
of the initial system can easily be recalculated in the new independent variables.   Therefore, reciprocal transformations preserve the whole class of integrable systems of hydrodynamic type.

The only object which, under reciprocal transformations,  behaves in a non-trivial way,  is the Hamiltonian structure. Generic reciprocal transformations destroy the local Hamiltonian formalism of the form (\ref{ham}). To the best of our knowledge, the behaviour of Hamiltonian structures under
reciprocal transformations (an their appropriate generalizations) has not been discussed in the literature. Investigation of this problem is, thus, the main objective of our paper.

In sections 3 and 4 we study the behaviour of Hamiltonian structures (%
\ref{ham}) under reciprocal transformations (\ref{recip}). The main
conclusion is that local Hamiltonian structures generally become nonlocal,
so that the transformed system (\ref{modif}) is still Hamiltonian, $u_{%
\tilde{t}}^{i}=\tilde{J}^{ij}\partial \tilde{h}/\partial u^{j}$, however,
with the nonlocal Hamiltonian operator%
\begin{equation}
\tilde{J}^{ij}=\tilde{g}^{ij}\frac{d}{d\tilde{x}}-\tilde{g}^{is}\tilde{\Gamma}_{sk}^{j}
u_{\tilde{x}}^{k}+\sum \varepsilon ^{\alpha }w_{(\alpha )k}^{i}u_{
\tilde{x}}^{k}\left( \frac{d}{d\tilde{x}}\right) ^{-1}w_{(\alpha )s}^{j}u_{
\tilde{x}}^{s}\text{. \ \ \ }  \label{nonloc}
\end{equation}%
Here $\varepsilon ^{\alpha }=\pm 1$; the explicit form of the transformed
metric $\tilde{g}^{ij}$ and the nonlocal terms is stated in sections 3 and
4.
 
\textbf{Remark.} In the particular case of constant $A$, $B$, $M$, $N$,
transformations (\ref{recip}) reduce to linear changes of the independent
variables, $\tilde{x}=bx+at$, $\tilde{t}=nx+mt$, ($a$, $b$, $m$, $n$ are
constants). As pointed out by Tsarev \cite{ts91}, transformations of this
type preserve the locality of Hamiltonian structures of hydrodynamic type
(see also \cite{max94}).
 
Nonlocal operators of the form (\ref{nonloc}) have been extensively
investigated in \cite{mh}, \cite{fr95}, \cite{fr95a}, \cite{max02}, \cite%
{al94}, \cite{ml01}, \cite{bg}, \cite{al95}, and have a remarkable
differential-geometric interpretation. The general theory is briefly
recalled in section 2. As an illustration of our approach, we discuss
Hamiltonian formalism of linearly degenerate semi-Hamiltonian systems in
Riemann invariants,
 
\begin{equation*}
R_{t}^{i}=\upsilon ^{i}(R)R_{x}^{i},
\end{equation*}%
where the characteristic speeds $\upsilon ^{i}(R)$ satisfy the identities
 
\bigskip

$1$.\ \ \ $\partial _{i}\upsilon ^{i}=0$ \ \ \ for any \ $i=1,2,...n$, $%
\partial _{i}=\partial /\partial R^{i}$ \ \ \ \ \ \ \ \ (linear
degeneracy);

\bigskip
 
$2$. \ \ $\partial _{k}\frac{\partial _{j}\upsilon ^{i}}{\upsilon
^{j}-\upsilon ^{i}}=\partial _{j}\frac{\partial _{k}\upsilon ^{i}}{\upsilon
^{k}-\upsilon ^{i}}$ \ \ \ \ \ for any$\ i\neq j\neq k$ \ \
(semi--Hamiltonian property).

\bigskip
 
\noindent Linearly degenerate semi-Hamiltonian systems have interesting interrelations
with separable (St\"{a}ckel) systems in classical mechanics \cite{ff1}, \cite%
{ff2}, \cite{bl}, \cite{fr91a}, \cite{fr92} and finite-gap solutions of 
integrable soliton equations. For instance, the linearly degenerate
semi-Hamiltonian system $R_{t}^{i}=\left( \sum R^{m}-R^{i}\right) R_{x}^{i}$
governs the evolution of zeros of the $n-$gap $\psi -$function in the KdV
theory \cite{dn75}, \cite{dn75a}, \cite{fr91a}. For $n=2$ we have
essentially one nontrivial linearly degenerate system,%
\begin{equation}
u_{t}=\upsilon u_{x}\text{, \ \ }\upsilon _{t}=u\upsilon _{x},  \label{2x2}
\end{equation}%
arising in gas dynamics (Chaplygin gas), field theory (Born-Infeld equation)
and classical differential geometry (minimal surfaces in Minkowski $3-$%
space, improper affine spheres). Theorem 1 of section 3 provides infinitely
many Hamiltonian representations of system (\ref{2x2}),%
\begin{equation}
\left(
\begin{array}{c}
u \\
\upsilon
\end{array}%
\right) _{t}=J\left(
\begin{array}{c}
\partial h/\partial u \\
\partial h/\partial \upsilon
\end{array}%
\right) ,  \label{hamoper}
\end{equation}%
with nonlocal operators%
\begin{equation*}
J=(u-\upsilon )^{2}\left(
\begin{array}{cc}
f(u) & 0 \\
0 & \varphi (\upsilon )%
\end{array}%
\right) \frac{d}{dx}+
\end{equation*}%
\begin{equation*}
(u-\upsilon )\left(
\begin{array}{cc}
\frac{1}{2}(u-\upsilon )f^{\prime }(u)u_{x}+f(u)(u_{x}-\upsilon _{x}) &
\varphi (\upsilon )u_{x}+f(u)\upsilon _{x} \\
-\varphi (\upsilon )u_{x}-f(u)\upsilon _{x} & \frac{1}{2}(u-\upsilon
)\varphi ^{\prime }(\upsilon )\upsilon _{x}+\varphi (\upsilon
)(u_{x}-\upsilon _{x})%
\end{array}%
\right) +
\end{equation*}%
\begin{equation*}
\left(
\begin{array}{cc}
w^{1}u_{x} & u_{x} \\
w^{2}\upsilon _{x} & \upsilon _{x}%
\end{array}%
\right) \left( \frac{d}{dx}\right) ^{-1}\left(
\begin{array}{cc}
u_{x} & \upsilon _{x} \\
w^{1}u_{x} & w^{2}\upsilon _{x}%
\end{array}%
\right),
\end{equation*}%
where $w^{1}=\frac{1}{2}[(u-\upsilon )f^{\prime }(u)-f(u)-\varphi (\upsilon )]
$ and $w^{2}=\frac{1}{2}[(\upsilon -u)\varphi ^{\prime }(\upsilon
)-f(u)-\varphi (\upsilon )]$ are the characteristic speeds of commuting
flows of system (\ref{2x2}), and the corresponding Hamiltonian densities%
\begin{equation*}
h=\frac{1}{4(\upsilon -u)}\left( \left[ \overset{u}{\int }\frac{d\nu }{\sqrt{%
f(\nu )}}\right] ^{2}-\left[ \overset{\upsilon }{\int }\frac{d\nu }{\sqrt{%
\varphi (\nu )}}\right] ^{2}\right) ,
\end{equation*}%
parametrized by two arbitrary functions $f(u)$ and $\varphi (\upsilon )$. In
view of the linearity of $J$ in $f(u)$ and $\varphi (\upsilon )$, all these
Hamiltonian structures are mutually compatible. Notice that the local part
of the operator $J$ is of the form (\ref{ham}), corresponding to the metric%
\begin{equation*}
(u-\upsilon )^{-2}\left( \frac{du^{2}}{f(u)}+\frac{d\upsilon ^{2}}{\varphi
(\upsilon )}\right) ,
\end{equation*}%
which is not flat for generic $f(u)$ and $\varphi (\upsilon )$. Particular
choices of $f(u)$ and $\varphi (\upsilon )$ lead to local Hamiltonian
structures which were first discovered in \cite{max87} and \cite{ar}
(see Example 1 in section 3).
 
Another example of this type is the $3-$component linearly degenerate
semi-Hamiltonian system%
\begin{equation}
u_{t}=(\upsilon +w)u_{x}\text{, \ \ }\upsilon _{t}=(u-w)\upsilon _{x}\text{,
\ \ \ }w_{t}=\frac{1}{2}(u+\upsilon )w_{x},  \label{sl}
\end{equation}%
which is the Riemann invariant form of non-isentropic gas dynamics with the
special equation of state $P(\rho ,s)=s-1/\rho $, see \cite{ver} and \cite%
{msh}. Our approach provides infinitely many nonlocal Hamiltonian structures
of the form%
\begin{eqnarray*}
&&
J=\frac{4}{\rho ^{2}}\left(
\begin{array}{ccc}
f(u) & 0 & 0 \\
0 & \varphi (\upsilon ) & 0 \\
0 & 0 & \psi (w)%
\end{array}%
\right) \frac{d}{dx}+ \\
&& \\
&&\frac{2}{\rho }\left(
\begin{array}{ccc}
\frac{1}{\rho }f^{\prime }(u)u_{x}+2f(u)(\frac{1}{\rho })_{x} & -\varphi
(\upsilon )u_{x}-f(u)\upsilon _{x} & -2\psi (w)u_{x}-f(u)w_{x} \\
\varphi (\upsilon )u_{x}+f(u)\upsilon _{x} & \frac{1}{\rho }\varphi ^{\prime
}(\upsilon )\upsilon _{x}+2\varphi (\upsilon )(\frac{1}{\rho })_{x} & -2\psi
(w)\upsilon _{x}+\varphi (\upsilon )w_{x} \\
2\psi (w)u_{x}+f(u)w_{x} & 2\psi (w)\upsilon _{x}-\varphi (\upsilon )w_{x} &
\frac{1}{\rho }\psi ^{\prime }(w)\upsilon _{x}+2\psi (w)(\frac{1}{\rho })_{x}%
\end{array}%
\right) + \\
&& \\
&&\left(
\begin{array}{cc}
w^{1}u_{x} & u_{x} \\
w^{2}\upsilon _{x} & \upsilon _{x} \\
w^{3}w_{x} & w_{x}%
\end{array}%
\right) \left( \frac{d}{dx}\right) ^{-1}\left(
\begin{array}{ccc}
u_{x} & \upsilon _{x} & w_{x} \\
w^{1}u_{x} & w^{2}\upsilon _{x} & w^{3}w_{x}%
\end{array}%
\right) ,\text{ \ \ }\rho =\left( w+\frac{\upsilon -u}{2}\right) ^{-1}.
\end{eqnarray*}%
Here the local part is of the form (\ref{ham}), generated by the diagonal
metric%
\begin{equation}
(2w+\upsilon -u)^{-2}\left( \frac{du^{2}}{f(u)}+\frac{d\upsilon ^{2}}{%
\varphi (\upsilon )}+\frac{dw^{2}}{\psi (w)}\right) ,  \label{met}
\end{equation}%
which is not flat in general, and $w^{1}$, $w^{2}$, $w^{3}$ are the
characteristic speeds of commuting flows of system (\ref{sl}):%
\begin{eqnarray*}
w^{1} &=&-\frac{1}{2}(2w+\upsilon -u)f^{\prime }(u)-\frac{1}{2}[f(u)+\varphi
(\upsilon )+4\psi (w)], \\
w^{2} &=&\frac{1}{2}(2w+\upsilon -u)\varphi ^{\prime }(\upsilon )-\frac{1}{2}%
[f(u)+\varphi (\upsilon )+4\psi (w)], \\
w^{3} &=&(2w+\upsilon -u)\psi ^{\prime }(w)-\frac{1}{2}[f(u)+\varphi
(\upsilon )+4\psi (w)].
\end{eqnarray*}%
These Hamiltonian operators depend linearly on three arbitrary functions $f(u)$, $%
\varphi (\upsilon )$ and $\psi (w)$, and are mutually compatible (Example 2
in section 3).
 
Similar arguments applied to the $3$-component linearly degenerate
semi-Hamiltonian system%
\begin{equation}
u_{t}=(\upsilon +w)u_{x}\text{, \ \ }\upsilon _{t}=(u+w)\upsilon _{x}\text{,
\ \ \ }w_{t}=(u+\upsilon )w_{x}  \label{3x3}
\end{equation}%
provide infinitely many nonlocal Hamiltonian structures%
\begin{equation*}
J=\left(
\begin{array}{ccc}
g^{11} & 0 & 0 \\
0 & g^{22} & 0 \\
0 & 0 & g^{33}%
\end{array}%
\right) \frac{d}{dx}+
\end{equation*}%
\begin{equation*}
\left(
\begin{array}{ccc}
g^{11}[\frac{f^{\prime }(u)}{2f(u)}u_{x}+\frac{u_{x}-\upsilon _{x}}{%
u-\upsilon }+\frac{u_{x}-w_{x}}{u-w}] & \frac{g^{22}u_{x}+g^{11}\upsilon _{x}%
}{u-\upsilon } & \frac{g^{33}u_{x}+g^{11}w_{x}}{u-w} \\
-\frac{g^{22}u_{x}+g^{11}\upsilon _{x}}{u-\upsilon } & g^{22}[\frac{\varphi
^{\prime }(\upsilon )}{2\varphi (\upsilon )}\upsilon _{x}+\frac{%
u_{x}-\upsilon _{x}}{u-\upsilon }+\frac{\upsilon _{x}-w_{x}}{\upsilon -w}] &
\frac{g^{33}\upsilon _{x}+g^{22}w_{x}}{\upsilon -w} \\
-\frac{g^{33}u_{x}+g^{11}w_{x}}{u-w} & -\frac{g^{33}\upsilon _{x}+g^{22}w_{x}%
}{\upsilon -w} & g^{33}[\frac{\psi ^{\prime }(w)}{2\psi (w)}w_{x}+\frac{%
w_{x}-\upsilon _{x}}{w-\upsilon }+\frac{u_{x}-w_{x}}{u-w}]%
\end{array}%
\right) +
\end{equation*}%
\begin{equation*}
\left(
\begin{array}{cccc}
\lambda ^{1}u_{x} & \mu ^{1}u_{x} & \eta ^{1}u_{x} & u_{x} \\
\lambda ^{2}\upsilon _{x} & \mu ^{2}\upsilon _{x} & \eta ^{2}\upsilon _{x} &
\upsilon _{x} \\
\lambda ^{3}w_{x} & \mu ^{3}w_{x} & \eta ^{3}w_{x} & w_{x}%
\end{array}%
\right) \left( \frac{d}{dx}\right) ^{-1}\left(
\begin{array}{ccc}
u_{x} & \upsilon _{x} & w_{x} \\
\eta ^{1}u_{x} & \eta ^{2}\upsilon _{x} & \eta ^{3}w_{x} \\
\mu ^{1}u_{x} & \mu ^{2}\upsilon _{x} & \mu ^{3}w_{x} \\
\lambda ^{1}u_{x} & \lambda ^{2}\upsilon _{x} & \lambda ^{3}w_{x}%
\end{array}%
\right) .
\end{equation*}%
Here the metric components are%
\begin{equation*}
g^{11}=(u-\upsilon )^{2}(u-w)^{2}f(u)\text{, \ \ }g^{22}=(\upsilon
-u)^{2}(\upsilon -w)^{2}\varphi (\upsilon )\text{, \ \ }g^{33}=(w-u)^{2}(w-%
\upsilon )^{2}\psi (w),
\end{equation*}%
the coefficients $\eta ^{1}=\upsilon +w$, $\eta ^{2}=u+w$ and $\eta
^{3}=\upsilon +w$ are the characteristic speeds of system (\ref{3x3}), and $%
\mu ^{k}$, $\lambda ^{k}$ are the characteristic speeds of its commuting
flows:%
\begin{eqnarray*}
\mu ^{1} &=&-\frac{1}{2}(u-\upsilon )(u-w)f^{\prime }(u)+uf(u)+\upsilon
\varphi (\upsilon )+w\psi (w) \\
&&-\overset{u}{\int }f(\nu )d\nu -\overset{\upsilon }{\int }\varphi (\nu
)d\nu -\overset{w}{\int }\psi (\nu )d\nu -\frac{1}{2}(\upsilon +w)\left(
f(u)+\varphi (\upsilon )+\psi (w)\right) , \\
&& \\
\lambda ^{1} &=&(u-\upsilon )(u-w)[2f(u)+uf^{\prime
}(u)]-2[u^{2}f(u)+\upsilon ^{2}\varphi (\upsilon )+w^{2}\psi (w)]+ \\
&&+(\upsilon +w)[uf(u)+\upsilon \varphi (\upsilon )+w\psi (w)+\overset{u}{%
\int }f(\nu )d\nu +\overset{\upsilon }{\int }\varphi (\nu )d\nu +\overset{w}{%
\int }\psi (\nu )d\nu ].
\end{eqnarray*}%
($\mu ^{2}$,  $\mu ^{3}$ and $\lambda ^{2}$, $\lambda ^{3}$ can be obtained by a
cyclic permutation of $u$, $\upsilon $, $w$ and $f$, $\varphi$, $\psi$). 
These operators depend
linearly on three arbitrary functions $f(u)$, $\varphi (\upsilon )$ and $\psi
(w)$ and are mutually compatible. The details can be found in Example 3 in
section 4.
 
Generalized reciprocal transformations and their action on the local Poisson
brackets of Dubrovin-Novikov type are discussed in section 5.
 
\section{Nonlocal Hamiltonian operators of hydrodynamic type. Nonlocal
Hamiltonian formalism for semi-Hamiltonian systems}
 
In this section we recall the necessary information about nonlocal
Hamiltonian operators of hydrodynamic type%
\begin{equation}
J^{ij}=g^{ij}\frac{d}{dx}-g^{is}\Gamma _{sk}^{j}u_{x}^{k}+\underset{\alpha }{%
\sum }\varepsilon ^{\alpha }w_{(\alpha )k}^{i}u_{x}^{k}\left( \frac{d}{dx}%
\right) ^{-1}w_{(\alpha )n}^{i}u_{x}^{n}\text{, \ \ \ \ \ }\varepsilon
^{\alpha }=\pm 1,  \label{a}
\end{equation}%
which are defined by a metric $g^{ij}$ ($g^{ij}=g^{ji}$, $\det g^{ij}\neq 0$%
), the corresponding Levi-Civita connection $\Gamma _{sk}^{j}$ and a set of
affinors $w_{(\alpha )}$ forming the nonlocal ''tail''.
 
\begin{theorem}
\cite{fr95} The operator $J^{ij}$ is Hamiltonian if and only if the affinors
$w_{(\alpha )}$ satisfy the equations%
\begin{eqnarray}
g_{ik}w_{(\alpha )j}^{k} &=&g_{jk}w_{(\alpha )i}^{k},  \label{ab} \\
\nabla _{k}w_{(\alpha )j}^{i} &=&\nabla _{j}w_{(\alpha )k}^{i},  \label{ac}
\end{eqnarray}%
and the curvature tensor of the metric $g^{ij}$ has the expansion%
\begin{equation}
R_{kl}^{ij}=\underset{\alpha }{\sum }\varepsilon ^{\alpha }\left( w_{(\alpha
)k}^{i}w_{(\alpha )l}^{j}-w_{(\alpha )l}^{i}w_{(\alpha )k}^{j}\right) .
\label{ae}
\end{equation}%
Moreover, the set of affinors $w_{(\alpha )}$ must be commutative,%
\begin{equation}
\lbrack w_{(\alpha )}\text{, }w_{(\beta )}]=0.  \label{af}
\end{equation}
\end{theorem}
 
As pointed out in \cite{fr95}, equations (\ref{ab})--(\ref{af}) constitute
the Gauss-Codazzi-Ricci equations of submanifolds of pseudo-Euclidean spaces
with flat normal bundle. There are three particularly interesting special
cases of the general formula (\ref{a}).
 
\textbf{Hamiltonian operators associated with \textit{constant curvature}
metrics}, first introduced in \cite{mh} (see also \cite{max02}), are of the
form%
\begin{equation}
J^{ij}=g^{ij}\frac{d}{dx}-g^{is}\Gamma _{sk}^{j}u_{x}^{k}+cu_{x}^{i}\left(
\frac{d}{dx}\right) ^{-1}u_{x}^{j}.  \label{ag}
\end{equation}%
Here $g^{ij}$ is a metric of constant curvature $c$, so that the Gauss
equation (\ref{ae}) takes the form
\begin{equation*}
R_{kl}^{ij}=c\left( \delta _{k}^{i}\delta _{l}^{j}-\delta _{l}^{i}\delta
_{k}^{j}\right) .
\end{equation*}
 
\textbf{Hamiltonian operators associated with \textit{conformally flat}
metrics} are of the form \cite{fr95a}%
\begin{equation}
J^{ij}=g^{ij}\frac{d}{dx}-g^{is}\Gamma
_{sk}^{j}u_{x}^{k}+w_{k}^{i}u_{x}^{k}\left( \frac{d}{dx}\right)
^{-1}u_{x}^{j}+u_{x}^{i}\left( \frac{d}{dx}\right) ^{-1}w_{k}^{j}u_{x}^{k}.
\label{ah}
\end{equation}%
The corresponding equations (\ref{ab})--(\ref{ae}) reduce to%
\begin{eqnarray*}
g_{ik}w_{j}^{k} &=&g_{jk}w_{i}^{k}, \\
\nabla _{k}w_{j}^{i} &=&\nabla _{j}w_{k}^{i},
\end{eqnarray*}%
while the curvature tensor of the metric $g^{ij}$ has the expansion%
\begin{equation*}
R_{kl}^{ij}=w_{k}^{i}\delta _{l}^{j}+w_{l}^{j}\delta
_{k}^{i}-w_{k}^{j}\delta _{l}^{i}-w_{l}^{i}\delta _{k}^{j},
\end{equation*}%
implying that the metric $g^{ij}$ is conformally flat.
 
\textbf{Hamiltonian operators associated with \textit{hypersurfaces of the
Euclidean space} }are \cite{fr95}%
\begin{equation}
J^{ij}=g^{ij}\frac{d}{dx}-g^{is}\Gamma
_{sk}^{j}u_{x}^{k}+w_{k}^{i}u_{x}^{k}\left( \frac{d}{dx}\right)
^{-1}w_{n}^{j}u_{x}^{n}.  \label{aj}
\end{equation}%
The corresponding equations (\ref{ab})--(\ref{ae}),%
\begin{equation*}
g_{ik}w_{j}^{k}=g_{jk}w_{i}^{k},
\end{equation*}%
\begin{equation*}
\nabla _{k}w_{j}^{i}=\nabla _{j}w_{k}^{i},
\end{equation*}%
\begin{equation*}
R_{kl}^{ij}=w_{k}^{i}w_{l}^{j}-w_{l}^{i}w_{k}^{j},
\end{equation*}%
are the Gauss-Codazzi equations governing the first fundamental form $g^{ij}$
and the shape operator $w_{j}^{i}$ of a hypersurface in the Euclidean space.
 
Nonlocal Hamiltonian operators of the form (\ref{a}) and their particular
instances (\ref{ag})--(\ref{aj}) appear in a variety of applications
including the Whitham theory (where operators of the form (\ref{a}) arise as
a result of the Whitham averaging of nonlocal higher order Poisson brackets of
integrable PDE's \cite{al94}, \cite{al95}), recursion scheme, Dirac
reduction, etc \cite{fr95}. Canonical coordinates for these operators were
introduced in a recent paper by Maltsev and Novikov \cite{ml01}. The
corresponding nonlocal symplectic structures were investigated in \cite{ml02}%
.
 
Notice that equations (\ref{a})--(\ref{ae}) simplify if the metric $g$ and
affinors $w_{(\alpha )}$ are diagonal, $g^{ij}=g^{ii}\delta ^{ij}$, $%
w_{(\alpha )j}^{i}=w_{(\alpha )}^{i}\delta _{j}^{i}$. In this case the
operator (\ref{a}) takes the form%
\begin{equation}
J^{ij}=g^{ii}\delta ^{ij}\frac{d}{dx}-g^{ii}\Gamma _{ik}^{j}u_{x}^{k}+%
\underset{\alpha }{\sum }\varepsilon ^{(\alpha )}w_{(\alpha
)}^{i}u_{x}^{i}\left( \frac{d}{dx}\right) ^{-1}w_{(\alpha )}^{j}u_{x}^{j},
\label{ai}
\end{equation}%
while equations (\ref{ab})--(\ref{ae}) reduce to%
\begin{equation}
\partial _{j}\ln \sqrt{g_{ii}}=\frac{\partial _{j}w_{(\alpha )}^{i}}{%
w_{(\alpha )}^{j}-w_{(\alpha )}^{i}},  \label{q}
\end{equation}%
\begin{equation}
R_{ij}^{ij}=\underset{\alpha }{\sum }\varepsilon ^{(\alpha )}w_{(\alpha
)}^{i}w_{(\alpha )}^{j}  \label{p}
\end{equation}%
for any\ $i\neq j$. Nonlocal Hamiltonian operators of the form (\ref{ai})
naturally arise as Hamiltonian structures of semi-Hamiltonian systems in
Riemann invariants \cite{fr95}, \cite{bg}. Let us recall the main
construction. Following Tsarev \cite{ts91}, a system in Riemann invariants%
\begin{equation}
R_{t}^{i}=\upsilon ^{i}(R)R_{x}^{i}\text{, \ \ \ }i=1,2,..., n  \label{ak}
\end{equation}%
is called semi-Hamiltonian if the characteristic speeds $\upsilon ^{i}(R)$
satisfy the identities%
\begin{equation*}
\partial _{k}\frac{\partial _{j}\upsilon ^{i}}{\upsilon ^{j}-\upsilon ^{i}}%
=\partial _{j}\frac{\partial _{k}\upsilon ^{i}}{\upsilon ^{k}-\upsilon ^{i}}
\end{equation*}%
for any triple $i\neq j\neq k$. Let us introduce the diagonal metric $%
g_{ii}dR^{i^{2}}$ by the formulae%
\begin{equation}
\partial _{j}\ln \sqrt{g_{ii}}=\frac{\partial _{j}\upsilon ^{i}}{\upsilon
^{j}-\upsilon ^{i}}  \label{r}
\end{equation}%
for any\ $i\neq j$. Notice that $g_{ii}$ is defined up to a multiple $%
g_{ii}\rightarrow g_{ii}/f^{i}(R^{i})$, where $f^{i}(R^{i})$ is an arbitrary
function of $R^{i}$.
 
Although the metric $g_{ii}$ may happen to be flat for particular choices of
$f^{i}(R^{i})$ (in this case our system is Hamiltonian with the local
Hamiltonian operator of  Dubrovin-Novikov type corresponding to the
metric $g_{ii}$), it is not flat in general. Suppose one can find the
expansion of the curvature tensor of this metric in the form (\ref{p}),
where $w_{(\alpha )}^{i}$ are characteristic speeds of commuting flows of
system (\ref{ak}):%
\begin{equation}
R_{t^{\alpha }}^{i}=w_{(\alpha )}^{i}(R)R_{x}^{i}.  \label{al}
\end{equation}%
Notice that the flows (\ref{ak}) and (\ref{al}) commute if and only if%
\begin{equation}
\frac{\partial _{j}\upsilon ^{i}}{\upsilon ^{j}-\upsilon ^{i}}=\frac{%
\partial _{j}w_{(\alpha )}^{i}}{w_{(\alpha )}^{j}-w_{(\alpha )}^{i}}
\label{oma}
\end{equation}%
for any\ $i\neq j$. In this case the system (\ref{ak}) will be Hamiltonian with the nonlocal
Hamiltonian operator (\ref{ai}) (indeed, both conditions (\ref{q}) and (\ref%
{p}) are satisfied). The main problem is thus to find the expansion (\ref%
{p}) for a metric satisfying (\ref{r}). We point out that the sum in (\ref{p}%
) is infinite in general.
 
Remarkably, there exist semi-Hamiltonian systems for which one can
explicitly construct the expansion (\ref{p}) for an arbitrary metric $g_{ii}$
satisfying (\ref{r}). Moreover, for any choice of  $f^{i}(R^{i})
$ the summation in (\ref{p}) is \textit{finite}, so that the corresponding
systems possess infinitely many Hamiltonian structures explicitly
parametrized by $n$ arbitrary functions of one variable. Among the most
interesting examples of this type are systems of Temple's class and linearly
degenerate semi-Hamiltonian systems. The construction of nonlocal
Hamiltonian formalism for linearly degenerate semi-Hamiltonian systems is
based on the formulae for reciprocal transformations of local Hamiltonian
operators which we derive in sections 3-5.
 
\section{Reciprocal transformations which change only the space variable $x$}
 
Consider a Hamiltonian system (\ref{hydro}),%
\begin{equation*}
u_{t}^{i}=\upsilon _{j}^{i}(u)u_{x}^{j}=(\nabla ^{i}\nabla
_{j}h)u_{x}^{j}=J^{ij}\frac{\partial h}{\partial u^{j}},
\end{equation*}%
with the local Hamiltonian operator (\ref{ham}),%
\begin{equation*}
J^{ij}=g^{ij}\frac{d}{dx}-g^{is}\Gamma _{sk}^{j}u_{x}^{k}.
\end{equation*}%
In this section we discuss special reciprocal transformations of the form (%
\ref{recip}) which change the space variable $x$ only,%
\begin{equation}
d\tilde{x}=B(u)dx+A(u)dt\text{, \ \ }\tilde{t}=t.  \label{recipx}
\end{equation}%
Notice that an arbitrary reciprocal transformation (\ref{recip}) can be
represented as a composition%
\begin{equation*}
R_{1}\circ T\circ R_{2}
\end{equation*}%
where $R_{1}$ and $R_{2}$ are reciprocal transformations of the form (\ref%
{recipx}) and $T$ is the transformation interchanging the independent
variables: $\tilde{x}=t$, $\tilde{t}=x$. After the transformation (\ref%
{recipx}), system (\ref{hydro}) takes the form%
\begin{equation}
u_{t}^{i}=\tilde{\upsilon}_{j}^{i}(u)u_{\tilde{x}}^{j}=\left( \nabla
^{i}\nabla _{j}h\cdot B-\delta _{j}^{i}A\right) u_{\tilde{x}}^{j}
\label{hydromod}
\end{equation}%
while the Hamiltonian operator $J^{ij}$ undergoes a nontrivial
transformation and becomes nonlocal. To write down the transformed operator $%
\tilde{J}^{ij}$, we introduce the Hamiltonian system%
\begin{equation}
u_{\tau }^{i}=w_{j}^{i}(u)u_{x}^{j}=J^{ij}\frac{\partial B}{\partial u^{j}}%
=(\nabla ^{i}\nabla _{j}B)u_{x}^{j}  \label{comm}
\end{equation}%
generated by the Hamiltonian operator $J^{ij}$ and the density $B$. Clearly,
systems (\ref{hydro}) and (\ref{comm}) commute. Since the flux of the
conserved density $B$ corresponding to the flow (\ref{comm}) is given by the
formula $\frac{1}{2}(\nabla B)^{2}$ (here $\nabla B=B_{i}du^{i}$ is the
gradient of $B$ and $(\nabla B)^{2}=g^{ij}B_{i}B_{j}$), we can incorporate
the time $\tau $ into the reciprocal transformation (\ref{recipx}) as
follows:%
\begin{equation*}
d\tilde{x}=Bdx+Adt+\frac{1}{2}(\nabla B)^{2}d\tau .
\end{equation*}%
Therefore, the transformed system (\ref{comm}) takes the form%
\begin{equation}
u_{\tau }^{i}=\tilde{w}_{j}^{i}(u)u_{\tilde{x}}^{j}=\left( \nabla ^{i}\nabla
_{j}B\cdot B-\frac{1}{2}\delta _{j}^{i}(\nabla B)^{2}\right) u_{\tilde{x}%
}^{j}  \label{commod}
\end{equation}%
The transformed systems (\ref{hydromod}) and (\ref{commod}) commute. Now we
can formulate the main result of this section.
 
\begin{theorem}
The transformed system (\ref{hydromod}) is Hamiltonian,%
\begin{equation*}
u_{t}^{i}=\tilde{J}^{ij}\frac{\partial \tilde{h}}{\partial u^{j}},
\end{equation*}%
with the nonlocal operator%
\begin{equation}
\tilde{J}^{ij}=\tilde{g}^{ij}\frac{d}{d\tilde{x}}-\tilde{g}^{is}\tilde{\Gamma%
}_{sk}^{j}u_{\tilde{x}}^{k}+\tilde{w}_{k}^{i}u_{\tilde{x}}^{k}\left( \frac{d%
}{d\tilde{x}}\right) ^{-1}u_{\tilde{x}}^{j}+u_{\tilde{x}}^{i}\left( \frac{d}{%
d\tilde{x}}\right) ^{-1}\tilde{w}_{k}^{j}u_{\tilde{x}}^{k}  \label{conf}
\end{equation}%
and the Hamiltonian density $\tilde{h}(u)=h(u)/B(u)$. Here the transformed
metric is $\tilde{g}^{ij}=B^{2}g^{ij}$, $\tilde{\Gamma}$ is the Levi-Civita
connection of $\tilde{g}$, and $\tilde{w}_{j}^{i}$ is given by (\ref{commod}%
). Notice that the transformed metric is conformally flat with the curvature
tensor%
\begin{equation*}
\tilde{R}_{kl}^{ij}=\tilde{w}_{k}^{i}\delta _{l}^{j}+\tilde{w}_{l}^{j}\delta
_{k}^{i}-\tilde{w}_{k}^{j}\delta _{l}^{i}-\tilde{w}_{l}^{i}\delta _{k}^{j}.
\end{equation*}
\end{theorem}
 
The proof follows directly from results of \cite{fr95a} where nonlocal
Hamiltonian operators of the form (\ref{conf}) associated with conformally
flat metrics were investigated. As an illustration of the procedure outlined
in Theorem 1, we explicitly construct nonlocal Hamiltonian formalism for the
$2-$component linearly degenerate system (\ref{2x2}).
 
\textbf{Example} \textbf{1}. The linear system%
\begin{equation}
u_{t}=u_{x}\text{, \ \ \ \ }\upsilon _{t}=-\upsilon _{x}  \label{lin}
\end{equation}%
possesses infinitely many Hamiltonian representations%
\begin{equation}
\left(
\begin{array}{c}
u \\
\upsilon
\end{array}%
\right) _{t}=J\left(
\begin{array}{c}
\partial h/\partial u \\
\partial h/\partial \upsilon
\end{array}%
\right)   \label{hamop}
\end{equation}%
with local Hamiltonian operators $J$ of the type (\ref{ham}),%
\begin{equation}
J=4\left(
\begin{array}{cc}
f(u) & 0 \\
0 & \varphi (\upsilon )%
\end{array}%
\right) \frac{d}{dx}+2\left(
\begin{array}{cc}
f^{\prime }(u)u_{x} & 0 \\
0 & \varphi ^{\prime }(\upsilon )\upsilon _{x}%
\end{array}%
\right) ,  \label{hamlin}
\end{equation}%
corresponding to flat metrics%
\begin{equation}
ds^{2}=\frac{du^{2}}{4f(u)}+\frac{d\upsilon ^{2}}{4\varphi (\upsilon )}
\label{metr}
\end{equation}%
(here $f(u)$ and $\varphi (\upsilon )$ are arbitrary functions). The
corresponding Hamiltonian densities are%
\begin{equation}
h(u,\upsilon )=\frac{1}{8}\left[ \overset{u}{\int }\frac{d\nu }{\sqrt{f(\nu )%
}}\right] ^{2}-\frac{1}{8}\left[ \overset{\upsilon }{\int }\frac{d\nu }{%
\sqrt{\varphi (\nu )}}\right] ^{2}.  \label{gam}
\end{equation}%
Consider the reciprocal transformation%
\begin{equation}
d\tilde{x}=Bdx+Adt=\frac{1}{2}(\upsilon -u)dx-\frac{1}{2}(\upsilon +u)dt%
\text{, \ \ \ \ \ }\tilde{t}=t,  \label{recipr}
\end{equation}%
mapping the linear system (\ref{lin}) to the linearly degenerate system%
\begin{equation}
u_{t}=\upsilon u_{\tilde{x}}\text{, \ \ \ }\upsilon _{t}=u\upsilon _{\tilde{x%
}}.  \label{lindeg}
\end{equation}%
To write down the transformed Hamiltonian operators (\ref{hamlin}), we first
introduce the transformed metric%
\begin{equation}
d\tilde{s}^{2}=\frac{ds^{2}}{B^{2}}=(u-\upsilon )^{-2}\left( \frac{du^{2}}{%
f(u)}+\frac{d\upsilon ^{2}}{\varphi (\upsilon )}\right) ,  \label{metrdef}
\end{equation}%
which is no longer flat for generic $f(u)$ and $\varphi (\upsilon )$.
Following the procedure outlined in Theorem 1, we calculate the Hamiltonian
flow generated by the Hamiltonian operator (\ref{hamlin}) and the
Hamiltonian density $B=\frac{1}{2}(\upsilon -u)$. A simple computation gives
the system%
\begin{equation}
u_{\tau }=-f^{\prime }(u)u_{x}\text{, \ \ \ }\upsilon _{\tau }=\varphi
^{\prime }(\upsilon )\upsilon _{x},  \label{lincompot}
\end{equation}%
which clearly commutes with (\ref{lin}). To incorporate the time $\tau $
into the reciprocal transformation (\ref{recipr}), we calculate the flux of
the density $B$ corresponding to the flow (\ref{lincompot}),%
\begin{equation*}
\frac{1}{2}(\nabla B)^{2}=\frac{1}{2}[f(u)+\varphi (\upsilon )].
\end{equation*}%
Thus, reciprocal transformation (\ref{recipr}) takes the form%
\begin{equation*}
d\tilde{x}=Bdx+Adt+\frac{1}{2}(\nabla B)^{2}d\tau =\frac{1}{2}(\upsilon
-u)dx-\frac{1}{2}(\upsilon +u)dt+\frac{1}{2}[f(u)+\varphi (\upsilon )]d\tau
\text{, \ \ }\tilde{t}=t,
\end{equation*}%
so that the transformed system (\ref{lincompot}) is%
\begin{equation}
u_{\tau }=w^{1}u_{\tilde{x}}\text{, \ \ \ }\upsilon _{\tau }=w^{2}\upsilon _{%
\tilde{x}},  \label{compot}
\end{equation}%
where%
\begin{equation}
w^{1}=\frac{1}{2}[(u-\upsilon )f^{\prime }(u)-f(u)-\varphi (\upsilon )]\text{%
, \ \ \ \ }w^{2}=\frac{1}{2}[(\upsilon -u)\varphi ^{\prime }(\upsilon
)-f(u)-\varphi (\upsilon )]  \label{w}
\end{equation}%
(notice that equations (\ref{compot}) and (\ref{w}) constitute a general
commuting flow of the linearly degenerate system (\ref{lindeg})). According
to Theorem 1, system (\ref{lindeg}) is Hamiltonian,%
\begin{equation*}
\left(
\begin{array}{c}
u \\
\upsilon
\end{array}%
\right) _{t}=\tilde{J}\left(
\begin{array}{c}
\partial \tilde{h}/\partial u \\
\partial \tilde{h}/\partial \upsilon
\end{array}%
\right) ,
\end{equation*}%
with nonlocal operators $\tilde{J}$ of the form%
\begin{equation*}
(u-\upsilon )^{2}\left(
\begin{array}{cc}
f(u) & 0 \\
0 & \varphi (\upsilon )%
\end{array}%
\right) \frac{d}{d\tilde{x}}+
\end{equation*}%
\begin{equation}
(u-\upsilon )\left(
\begin{array}{cc}
\frac{1}{2}(u-\upsilon )f^{\prime }(u)u_{\tilde{x}}+f(u)(u_{\tilde{x}%
}-\upsilon _{\tilde{x}}) & \varphi (\upsilon )u_{\tilde{x}}+f(u)\upsilon _{%
\tilde{x}} \\
-\varphi (\upsilon )u_{\tilde{x}}-f(u)\upsilon _{\tilde{x}} & \frac{1}{2}%
(u-\upsilon )\varphi ^{\prime }(\upsilon )\upsilon _{\tilde{x}}\text{+}%
\varphi (\upsilon )(u_{\tilde{x}}-\upsilon _{\tilde{x}})%
\end{array}%
\right)   \label{opt}
\end{equation}%
\begin{equation*}
+\left(
\begin{array}{cc}
w^{1}u_{\tilde{x}} & u_{\tilde{x}} \\
w^{2}\upsilon _{\tilde{x}} & \upsilon _{\tilde{x}}%
\end{array}%
\right) \left( \frac{d}{d\tilde{x}}\right) ^{-1}\left(
\begin{array}{cc}
u_{\tilde{x}} & \upsilon _{\tilde{x}} \\
w^{1}u_{\tilde{x}} & w^{2}\upsilon _{\tilde{x}}%
\end{array}%
\right)
\end{equation*}%
and the Hamiltonian densities%
\begin{equation*}
\tilde{h}=h/B=\frac{1}{4(\upsilon -u)}\left( \left[ \overset{u}{\int }\frac{%
d\nu }{\sqrt{f(\nu )}}\right] ^{2}-\left[ \overset{\upsilon }{\int }\frac{%
d\nu }{\sqrt{\varphi (\nu )}}\right] ^{2}\right) .
\end{equation*}%
All these operators are mutually compatible. Let us discuss some particular
cases. For $f(u)=1$, $\varphi (\upsilon )=-1$ and $f(u)=u$, $\varphi
(\upsilon )=-\upsilon $ we have $w^{1}=w^{2}=0$, so that the transformed
operators take local forms%
\begin{equation}
(u-\upsilon )^{2}\left(
\begin{array}{cc}
1 & 0 \\
0 & -1%
\end{array}%
\right) \frac{d}{d\tilde{x}}+(u-\upsilon )\left(
\begin{array}{cc}
u_{\tilde{x}}-\upsilon _{\tilde{x}} & \upsilon _{\tilde{x}}-u_{\tilde{x}} \\
u_{\tilde{x}}-\upsilon _{\tilde{x}} & \upsilon _{\tilde{x}}-u_{\tilde{x}}%
\end{array}%
\right)   \label{op1}
\end{equation}%
and%
\begin{equation}
(u-\upsilon )^{2}\left(
\begin{array}{cc}
u & 0 \\
0 & -\upsilon
\end{array}%
\right) \frac{d}{d\tilde{x}}\text{+}(u-\upsilon )\left(
\begin{array}{cc}
(\frac{3}{2}u-\frac{1}{2}\upsilon )u_{\tilde{x}}-u\upsilon _{\tilde{x}} &
u\upsilon _{\tilde{x}}-\upsilon u_{\tilde{x}} \\
\upsilon u_{\tilde{x}}-u\upsilon _{\tilde{x}} & (\frac{3}{2}\upsilon -\frac{1%
}{2}u)\upsilon _{\tilde{x}}-\upsilon u_{\tilde{x}}%
\end{array}%
\right) ,  \label{ops2}
\end{equation}%
the corresponding Hamiltonian densities being%
\begin{equation*}
h(u,\upsilon )=\frac{u^{2}+\upsilon ^{2}}{4(\upsilon -u)}\text{ \ \ \ \ and
\ \ \ \ }h(u,\upsilon )=\frac{u+\upsilon }{\upsilon -u},
\end{equation*}%
respectively. In the case $f(u)=u^{2}$, $\varphi (\upsilon )=-\upsilon ^{2}$
we have $w^{1}=-w^{2}=\frac{1}{2}(u-\upsilon )^{2}$ so that the operator $\tilde{J}$
takes the form%
\begin{eqnarray}
&&(u-\upsilon )^{2}\left(
\begin{array}{cc}
u^{2} & 0 \\
0 & -\upsilon ^{2}%
\end{array}%
\right) \frac{d}{d\tilde{x}}+  \notag \\
&&(u-\upsilon )\left(
\begin{array}{cc}
u(2u-\upsilon )u_{\tilde{x}}-u^{2}\upsilon _{\tilde{x}} & u^{2}\upsilon _{%
\tilde{x}}-\upsilon ^{2}u_{\tilde{x}} \\
\upsilon ^{2}u_{\tilde{x}}-u^{2}\upsilon _{\tilde{x}} & \upsilon (2\upsilon
-u)\upsilon _{\tilde{x}}-\upsilon ^{2}u_{\tilde{x}}%
\end{array}%
\right) +  \label{opa} \\
&&\left(
\begin{array}{cc}
\frac{1}{2}(u-\upsilon )^{2}u_{\tilde{x}} & u_{\tilde{x}} \\
-\frac{1}{2}(u-\upsilon )^{2}\upsilon _{\tilde{x}} & \upsilon _{\tilde{x}}%
\end{array}%
\right) \left( \frac{d}{d\tilde{x}}\right) ^{-1}\left(
\begin{array}{cc}
u_{\tilde{x}} & \upsilon _{\tilde{x}} \\
\frac{1}{2}(u-\upsilon )^{2}u_{\tilde{x}} & -\frac{1}{2}(u-\upsilon
)^{2}\upsilon _{\tilde{x}}%
\end{array}%
\right)   \notag
\end{eqnarray}%
with the corresponding density%
\begin{equation*}
h(u,\upsilon )=\frac{(\ln u)^{2}+(\ln \upsilon )^{2}}{4(\upsilon -u)}.
\end{equation*}%
Notice that the local part of the Hamiltonian operator (\ref{opa}),%
\begin{equation}
(u-\upsilon )^{2}\left(
\begin{array}{cc}
u^{2} & 0 \\
0 & -\upsilon ^{2}%
\end{array}%
\right) \frac{d}{d\tilde{x}}\text{+}(u-\upsilon )\left(
\begin{array}{cc}
u(2u-\upsilon )u_{\tilde{x}}-u^{2}\upsilon _{\tilde{x}} & u^{2}\upsilon _{%
\tilde{x}}-\upsilon ^{2}u_{\tilde{x}} \\
\upsilon ^{2}u_{\tilde{x}}-u^{2}\upsilon _{\tilde{x}} & \upsilon (2\upsilon
-u)\upsilon _{\tilde{x}}-\upsilon ^{2}u_{\tilde{x}}%
\end{array}%
\right) ,  \label{op4}
\end{equation}%
is itself Hamiltonian, indeed, the metric%
\begin{equation*}
(u-\upsilon )^{-2}\left( \frac{du^{2}}{u^{2}}-\frac{d\upsilon ^{2}}{\upsilon
^{2}}\right)
\end{equation*}%
is flat.
 
One can show that the operators (\ref{op1}), (\ref{ops2}) and (\ref{op4})
(are arbitrary linear combinations thereof) are the only local Hamiltonian
structures of Dubrovin-Novikov type of the system (\ref{lindeg}). This
follows from the fact that the metric%
\begin{equation*}
(u-\upsilon )^{-2}\left( \frac{du^{2}}{f(u)}-\frac{d\upsilon ^{2}}{\varphi
(\upsilon )}\right)
\end{equation*}%
is flat if and only if%
\begin{equation*}
f(u)=\alpha u^{2}+\beta u+\gamma \text{, \ \ \ }\varphi (\upsilon )=-\alpha
\upsilon ^{2}-\beta \upsilon -\gamma ,
\end{equation*}%
where $\alpha $, $\beta $ and $\gamma $ are arbitrary constants. These local
structures were first discovered in \cite{max87} and \cite{ar}, and
subsequently generalized to polytropic gas dynamics in \cite{nt}, \cite{ol}.
The general nonlocal operator (\ref{opt}) first appeared in \cite{bg}.
 
\textbf{Example 2}. The linear system%
\begin{equation}
u_{t}=u_{x}\text{, \ \ \ }\upsilon _{t}=-\upsilon _{x}\text{, \ \ \ }w_{t}=0
\label{bb}
\end{equation}%
possesses infinitely many Hamiltonian representations%
\begin{equation*}
\left(
\begin{array}{c}
u \\
\upsilon  \\
w%
\end{array}%
\right) _{t}=J\left(
\begin{array}{c}
\partial h/\partial u \\
\partial h/\partial \upsilon  \\
\partial h/\partial w%
\end{array}%
\right)
\end{equation*}%
with local operators%
\begin{equation*}
J=4\left(
\begin{array}{ccc}
f(u) & 0 & 0 \\
0 & \varphi (\upsilon ) & 0 \\
0 & 0 & \psi (w)%
\end{array}%
\right) \frac{d}{dx}+2\left(
\begin{array}{ccc}
f^{\prime }(u)u_{x} & 0 & 0 \\
0 & \varphi ^{\prime }(\upsilon )\upsilon _{x} & 0 \\
0 & 0 & \psi ^{\prime }(w)w_{x}%
\end{array}%
\right)
\label{333}
\end{equation*}%
corresponding to flat metrics%
\begin{equation*}
ds^{2}=\frac{du^{2}}{4f(u)}+\frac{d\upsilon ^{2}}{4\varphi (\upsilon )}+%
\frac{dw^{2}}{4\psi (w)}
\end{equation*}%
(here $f(u)$, $\varphi (\upsilon )$ and $\psi (w)$ are arbitrary functions). The
corresponding Hamiltonian densities are%
\begin{equation*}
h(u,\upsilon )=\frac{1}{8}\left[ \overset{u}{\int }\frac{d\nu }{\sqrt{f(\nu )%
}}\right] ^{2}-\frac{1}{8}\left[ \overset{\upsilon }{\int }\frac{d\nu }{%
\sqrt{\varphi (\nu )}}\right] ^{2}.
\end{equation*}%
Consider the reciprocal transformation%
\begin{equation}
d\tilde{x}=Bdx+Adt=[w+\frac{1}{2}(\upsilon -u)]dx-\frac{1}{2}(u+\upsilon )dt%
\text{, \ \ \ \ \ }\tilde{t}=t  \label{bc}
\end{equation}%
mapping the system (\ref{bb}) to the linearly degenerate semi-Hamiltonian
system%
\begin{equation}
u_{t}=(\upsilon +w)u_{\tilde{x}}\text{, \ \ \ \ }\upsilon _{t}=(u-w)\upsilon
_{\tilde{x}}\text{, \ \ \ \ \ }w_{t}=\frac{1}{2}(u+\upsilon )w_{\tilde{x}}.
\label{bd}
\end{equation}%
Notice that under the change of variables
\begin{equation}
u=-c+s-1/\rho \text{, \ \ \ }\upsilon =-c-s+1/\rho \text{, \ \ \ }w=s
\label{sun}
\end{equation}%
system (\ref{bd}) takes the form of equations of gas dynamics
\begin{equation}
\rho _{t}+\partial _{\tilde{x}}(\rho c)=0\text{, \ \ \ }c_{t}+cc_{x}+\frac{1%
}{\rho }\partial _{\tilde{x}}P(\rho \text{, }s)=0\text{, \ \ \ }s_{t}+cs_{%
\tilde{x}}=0  \label{be}
\end{equation}%
with the special equation of state $P=s-1/\rho $ generalizing isentropic
Chaplygin gas (see \cite{ver} and \cite{msh}).
 
To calculate the transformed Hamiltonian operators (\ref{333}), we
introduce the conformal metric (\ref{met}),%
\begin{equation*}
d\tilde{s}^{2}=\frac{ds^{2}}{B^{2}}=(2w+\upsilon -u)^{-2}\left( \frac{du^{2}%
}{f(u)}+\frac{d\upsilon ^{2}}{\varphi (\upsilon )}+\frac{dw^{2}}{\psi (w)}%
\right) ,
\end{equation*}%
which is no longer flat in general. Following the recipe of Theorem 1, we
introduce the Hamiltonian flow generated by the Hamiltonian operator (\ref%
{333}) and the Hamiltonian density $B=w+\frac{1}{2}(\upsilon -u)$. A
simple computation gives the system
\begin{equation}
u_{\tau }=-f^{\prime }(u)u_{x}\text{, \ \ \ }\upsilon _{\tau }=\varphi
^{\prime }(\upsilon )\upsilon _{x}\text{, \ \ \ }w_{\tau }=2\psi ^{\prime
}(w)w_{x}  \label{zu}
\end{equation}%
which commutes with (\ref{bb}). To incorporate the time $\tau $ into the
reciprocal transformation (\ref{bc}), we calculate the flux of the density $B
$ corresponding to the flow (\ref{zu}),%
\begin{equation*}
\frac{1}{2}(\nabla B)^{2}=\frac{1}{2}[f(u)+\varphi (\upsilon )+4\psi (w)].
\end{equation*}%
Thus, transformation (\ref{bc}) takes the form%
\begin{equation*}
\begin{array}{c}
d\tilde{x}=Bdx\text{+}Adt\text{+}\frac{1}{2}(\nabla B)^{2}d\tau =[w\text{+}%
\frac{1}{2}(\upsilon -u)]dx-\frac{1}{2}(\upsilon \text{+}u)dt\text{+}\frac{1%
}{2}[f(u)\text{+}\varphi (\upsilon )\text{+}4\psi (w)]d\tau, \\
\ \\
\tilde{t}=t,
\end{array}
\end{equation*}%
so that the transformed flow (\ref{zu}) is%
\begin{equation}
u_{\tau }=w^{1}u_{\tilde{x}}\text{, \ \ \ }\upsilon _{\tau }=w^{2}\upsilon _{%
\tilde{x}}\text{, \ \ \ }w_{\tau }=w^{3}w_{\tilde{x}},  \label{zum}
\end{equation}%
where%
\begin{eqnarray}
w^{1} &=&-\frac{1}{2}(2w+\upsilon -u)f^{\prime }(u)-\frac{1}{2}[f(u)+\varphi
(\upsilon )+4\psi (w)]\text{, \ \ \ \ }  \notag \\
w^{2} &=&\frac{1}{2}(2w+\upsilon -u)\varphi ^{\prime }(\upsilon )-\frac{1}{2}%
[f(u)+\varphi (\upsilon )+4\psi (w)],  \label{coma} \\
w^{3} &=&(2w+\upsilon -u)\psi ^{\prime }(w)-\frac{1}{2}[f(u)+\varphi
(\upsilon )+4\psi (w)].  \notag
\end{eqnarray}%
Notice that equations (\ref{zum}) and (\ref{coma}) constitute a general
commuting flow of the linearly degenerate system (\ref{bd}). According to
Theorem 1, the system (\ref{bd}) is Hamiltonian,%
\begin{equation*}
\left(
\begin{array}{c}
u \\
\upsilon  \\
w%
\end{array}%
\right) _{t}=\tilde{J}\left(
\begin{array}{c}
\partial \tilde{h}/\partial u \\
\partial \tilde{h}/\partial \upsilon  \\
\partial \tilde{h}/\partial w%
\end{array}%
\right) ,
\end{equation*}%
with nonlocal operators $\tilde{J}$ of the form%
\begin{equation*}
\frac{4}{\rho ^{2}}\left(
\begin{array}{ccc}
f(u) & 0 & 0 \\
0 & \varphi (\upsilon ) & 0 \\
0 & 0 & \psi (w)%
\end{array}%
\right) \frac{d}{d\tilde{x}}+
\end{equation*}%
\begin{equation}
\frac{2}{\rho }\left(
\begin{array}{ccc}
\frac{1}{\rho }f^{\prime }(u)u_{\tilde{x}}+2f(u)(\frac{1}{\rho })_{\tilde{x}}
& -\varphi (\upsilon )u_{\tilde{x}}-f(u)\upsilon _{\tilde{x}} & -2\psi (w)u_{%
\tilde{x}}-f(u)w_{\tilde{x}} \\
\varphi (\upsilon )u_{\tilde{x}}+f(u)\upsilon _{\tilde{x}} & \frac{1}{\rho }%
\varphi ^{\prime }(\upsilon )\upsilon _{\tilde{x}}+2\varphi (\upsilon )(%
\frac{1}{\rho })_{\tilde{x}} & -2\psi (w)\upsilon _{\tilde{x}}+\varphi
(\upsilon )w_{\tilde{x}} \\
2\psi (w)u_{\tilde{x}}+f(u)w_{\tilde{x}} & 2\psi (w)\upsilon _{\tilde{x}%
}-\varphi (\upsilon )w_{\tilde{x}} & \frac{1}{\rho }\psi ^{\prime
}(w)\upsilon _{\tilde{x}}+2\psi (w)(\frac{1}{\rho })_{\tilde{x}}%
\end{array}%
\right) +  \label{hhu}
\end{equation}%
\begin{equation*}
\left(
\begin{array}{cc}
w^{1}u_{\tilde{x}} & u_{\tilde{x}} \\
w^{2}\upsilon _{\tilde{x}} & \upsilon _{\tilde{x}} \\
w^{3}w_{\tilde{x}} & w_{\tilde{x}}%
\end{array}%
\right) \left( \frac{d}{d\tilde{x}}\right) ^{-1}\left(
\begin{array}{ccc}
u_{\tilde{x}} & \upsilon _{\tilde{x}} & w_{\tilde{x}} \\
w^{1}u_{\tilde{x}} & w^{2}\upsilon _{\tilde{x}} & w^{3}w_{\tilde{x}}%
\end{array}%
\right)
\end{equation*}%
and the Hamiltonian densities%
\begin{equation*}
\tilde{h}=h/B=\frac{\rho }{8}\left( \left[ \overset{u}{\int }\frac{d\nu }{%
\sqrt{f(\nu )}}\right] ^{2}-\left[ \overset{\upsilon }{\int }\frac{d\nu }{%
\sqrt{\varphi (\nu )}}\right] ^{2}\right) ,
\end{equation*}%
where $2w+\upsilon -u=2/\rho $ (see (\ref{sun})). We point out that the
corresponding metric (\ref{met}) is flat if and only if%
\begin{equation*}
f(u)=\varepsilon u+\alpha \text{, \ \ }\varphi (\upsilon )=\varepsilon
\upsilon +\beta \text{, \ \ \ }\psi (w)=-\frac{\varepsilon }{2}w+\gamma ,
\end{equation*}%
where the constants $\alpha $, $\beta $, $\gamma $ and $\varepsilon $ are
constrained by $\alpha +\beta +4\gamma =0$. In this case $w^{1}=w^{2}=w^{3}=0$, so that the corresponding operators (\ref{hhu}) are local, thus providing $%
3$ non-equivalent Hamiltonian structures of Dubrovin-Novikov type.
 
\section{Reciprocal transformations which change both $x$ and $t$}
 
Consider again the Hamiltonian system (\ref{hydro}),
\begin{equation*}
u_{t}^{i}=\upsilon _{j}^{i}(u)u_{x}^{j}=(\nabla ^{i}\nabla
_{j}h)u_{x}^{j}=J^{ij}\frac{\partial h}{\partial u^{j}},
\end{equation*}%
with the local Hamiltonian operator
\begin{equation*}
J^{ij}=g^{ij}\frac{d}{dx}-g^{is}\Gamma _{sk}^{j}u_{x}^{k}
\end{equation*}%
corresponding to the flat metric $g^{ij}$. In this section we discuss the
behavior of Hamiltonian operators (\ref{ham}) under general reciprocal
transformations%
\begin{equation*}
d\tilde{x}=B(u)dx+A(u)dt\text{, \ \ \ \ }d\tilde{t}=N(u)dx+M(u)dt
\end{equation*}%
of the form (\ref{recip}) which map system (\ref{hydro}) to the system (\ref%
{modif}),%
\begin{equation*}
u_{\tilde{t}}^{i}=\tilde{\upsilon}_{j}^{i}(u)u_{\tilde{x}}^{j}\text{, \ \ \
\ \ \ \ \ \ }\tilde{\upsilon}=(B\upsilon -AE)(ME-N\upsilon )^{-1}\text{, \ \
\ \ }E=id.
\end{equation*}%
First of all, we introduce the transformed metric%
\begin{equation}
\tilde{g}^{ij}=\frac{(M\delta _{s}^{i}-N\upsilon _{s}^{i})(M\delta
_{r}^{j}-N\upsilon _{r}^{j})}{(BM-AN)^{2}}g^{sr}  \label{new}
\end{equation}%
which is no longer flat in general. This metric will generate the local part
of the transformed Hamiltonian operator $\tilde{J}$.
 
\textbf{Remark}. In the particular case of reciprocal transformations
changing only the variable $x$, we have $N=0$, $M=1$ so that formula (\ref%
{new}) reduces to the conformal transformation%
\begin{equation*}
\tilde{g}^{ij}=g^{ij}/B^{2}.
\end{equation*}%
Similarly, if we interchange the independent variables 
(that is, if $A=N=1$, $B=M=0$), we have%
\begin{equation*}
\tilde{g}^{ij}=\upsilon _{s}^{i}g^{sr}\upsilon _{r}^{j},
\end{equation*}%
which coincides with the formula proposed in \cite{ts91}. Notice that in the
second case the metric $\tilde{g}^{ij}$ is automatically flat, which means
that the local Hamiltonian formalism is invariant under the interchange of
independent variables.
 
To calculate the nonlocal ''tail'' of the transformed operator $\tilde{J}$,
we introduce the Hamiltonian flows%
\begin{equation}
u_{\tau }^{i}=(\nabla ^{i}\nabla _{j}N)u_{x}^{j}=J^{ij}\frac{\partial N}{%
\partial u^{j}}  \label{no}
\end{equation}%
and%
\begin{equation}
u_{\zeta }^{i}=(\nabla ^{i}\nabla _{j}B)u_{x}^{j}=J^{ij}\frac{\partial B}{%
\partial u^{j}},  \label{bo}
\end{equation}%
generated by the Hamiltonian operator $J^{ij}$ and the Hamiltonian densities
$N$ and $B$ which enter the reciprocal transformation (\ref{recip}).
Clearly, systems (\ref{no}) and (\ref{bo}) commute with (\ref{hydro}). To
incorporate the times $\tau $ and $\zeta $ into the reciprocal
transformation (\ref{recip}), we calculate the fluxes $P$, $Q$ and $R$, $S$
of the Hamiltonian densities $B$ and $N$,%
\begin{eqnarray*}
d\tilde{x} &=&Bdx+Adt+Pd\tau +Qd\zeta , \\
d\tilde{t} &=&Ndx+Mdt+Rd\tau +Sd\zeta ,
\end{eqnarray*}%
that is, the coefficients which make these $1-$forms closed by virtue of (%
\ref{hydro}), (\ref{no}) and (\ref{bo}). Here%
\begin{eqnarray*}
Q &=&\frac{1}{2}(\nabla B)^{2}=\frac{1}{2}g^{ij}B_{i}B_{j}, \\
R &=&\frac{1}{2}(\nabla N)^{2}=\frac{1}{2}g^{ij}N_{i}N_{j},
\end{eqnarray*}%
(see (\cite{ts91})) and $P$, $S$ are normalized as%
\begin{equation*}
P+S=(\nabla B,\nabla N)=g^{ij}B_{i}N_{j}.
\end{equation*}%
Therefore, the transformed flows (\ref{no}) and (\ref{bo}) take the forms%
\begin{equation}
u_{\tau }^{i}=\mu _{j}^{i}(u)u_{\tilde{x}}^{j}  \label{newno}
\end{equation}%
and%
\begin{equation}
u_{\zeta }^{i}=\lambda _{j}^{i}(u)u_{\tilde{x}}^{j},  \label{newbo}
\end{equation}%
where the matrices $\mu _{j}^{i}$ and $\lambda _{j}^{i}$ are%
\begin{equation}
\mu _{j}^{i}=B\nabla ^{i}\nabla _{j}N-P\delta _{j}^{i}+N\tilde{\upsilon}%
_{j}^{k}\nabla ^{i}\nabla _{k}N-R\tilde{\upsilon}_{j}^{i}  \label{mu}
\end{equation}%
and%
\begin{equation}
\lambda _{j}^{i}=B\nabla ^{i}\nabla _{j}B-Q\delta _{j}^{i}+N\tilde{\upsilon}%
_{j}^{k}\nabla ^{i}\nabla _{k}B-S\tilde{\upsilon}_{j}^{i},  \label{lambda}
\end{equation}%
respectively (these formulae readily follow from (\ref{no}) and (\ref{bo})
after one changes from $x$, $t$, $\tau $, $\zeta $ to $\tilde{x}$, $\tilde{t}
$, $\tau $, $\zeta $). The transformed flows (\ref{newno}) and (\ref{newbo})
commute with (\ref{modif}). Now we can formulate the main result of this
section.
 
\begin{theorem}
After the reciprocal transformation (\ref{recip}), the local Hamiltonian
operator (\ref{ham}) corresponding to the flat metric $g^{ij}$ becomes
nonlocal of the form%
\begin{eqnarray*}
\tilde{J}^{ij} &=&\tilde{g}^{ij}\frac{d}{d\tilde{x}}-\tilde{g}^{is}\tilde{%
\Gamma}_{sk}^{j}u_{\tilde{x}}^{k}+\lambda _{k}^{i}u_{\tilde{x}}^{k}\left(
\frac{d}{d\tilde{x}}\right) ^{-1}u_{\tilde{x}}^{j}+u_{\tilde{x}}^{i}\left(
\frac{d}{d\tilde{x}}\right) ^{-1}\lambda _{k}^{j}u_{\tilde{x}}^{k} \\
&&+\mu _{k}^{i}u_{\tilde{x}}^{k}\left( \frac{d}{d\tilde{x}}\right) ^{-1}%
\tilde{\upsilon}_{n}^{j}u_{\tilde{x}}^{n}+\tilde{\upsilon}_{n}^{i}u_{\tilde{x%
}}^{n}\left( \frac{d}{d\tilde{x}}\right) ^{-1}\mu _{k}^{j}u_{\tilde{x}}^{k}.
\end{eqnarray*}%
Here the local part is defined by the metric $\tilde{g}^{ij}$ given by (\ref%
{new}), $\tilde{\Gamma}$ is the Levi-Civita connection of $\tilde{g}$, and
the nonlocal terms $\lambda _{j}^{i}$, $\mu _{j}^{i}$ and $\tilde{\upsilon}%
_{j}^{i}$ are defined in (\ref{mu}), (\ref{lambda}) and (\ref{modif}). In
particular, the curvature tensor of $\tilde{g}^{ij}$ is
\begin{equation*}
\tilde{R}_{kl}^{ij}=\lambda _{k}^{i}\delta _{l}^{j}+\lambda _{l}^{j}\delta
_{k}^{i}-\lambda _{k}^{j}\delta _{l}^{i}-\lambda _{l}^{i}\delta _{k}^{j}+\mu
_{k}^{i}\tilde{\upsilon}_{l}^{j}+\mu _{l}^{j}\tilde{\upsilon}_{k}^{i}-\mu
_{k}^{j}\tilde{\upsilon}_{l}^{i}-\mu _{l}^{i}\tilde{\upsilon}_{k}^{j}.
\end{equation*}
\end{theorem}
 
The proof of Theorem 3 will follow from general results of section 5.
 
As an illustration of this procedure, we explicitly calculate nonlocal
Hamiltonian operators associated with the $3-$component linearly degenerate
system (\ref{3x3}).
 
\textbf{Example} \textbf{3}. We start with the same linear system (\ref{bb})
as in Exampe 2,%
\begin{equation*}
u_{t}=u_{x},\ \ \ \ \upsilon _{t}=-\upsilon _{x},\ \ \ \ w_{t}=0
\end{equation*}%
which possesses infinitely many Hamiltonian structures with operators%
\begin{equation}
J=\left(
\begin{array}{ccc}
f(u) & 0 & 0 \\
0 & \varphi (\upsilon ) & 0 \\
0 & 0 & \frac{1}{4}\psi (w)%
\end{array}%
\right) \frac{d}{dx}+\frac{1}{2}\left(
\begin{array}{ccc}
f^{\prime }u_{x} & 0 & 0 \\
0 & \varphi ^{\prime }\upsilon _{x} & 0 \\
0 & 0 & \frac{1}{4}\psi ^{\prime }w_{x}%
\end{array}%
\right)   \label{s}
\end{equation}%
corresponding to flat metrics%
\begin{equation}
ds^{2}=\frac{du^{2}}{f(u)}+\frac{d\upsilon ^{2}}{\varphi (\upsilon )}+4\frac{%
dw^{2}}{\psi (w)},  \label{metr3x3}
\end{equation}%
(here $f(u)$, $\varphi (\upsilon )$ and $\psi (w)$ are arbitrary functions;
the multiple $4$ is chosen for the convenience).
 
Consider the reciprocal transformation%
\begin{eqnarray}
d\tilde{x} &=&Bdx+Adt=(2w^{2}-u^{2}-\upsilon ^{2})dx+(\upsilon ^{2}-u^{2})dt,
\label{10} \\
d\tilde{t} &=&Ndx+Mdt=(u+\upsilon -2w)dx+(u-\upsilon )dt,  \notag
\end{eqnarray}%
which maps system (\ref{bb}) to the $3-$component linearly degenerate
semi-Hamiltonian system%
\begin{equation}
u_{\tilde{t}}=(\upsilon +w)u_{\tilde{x}}\text{, \ \ \ }\upsilon _{\tilde{t}%
}=(u+w)\upsilon _{\tilde{x}}\text{, \ \ \ }w_{\tilde{t}}=(u+\upsilon )w_{%
\tilde{x}}.  \label{9}
\end{equation}%
The transformed metric (\ref{metr3x3}) (apply formula (\ref{new})),%
\begin{equation}
d\tilde{s}^{2}=\frac{du^{2}}{(u-\upsilon )^{2}(u-w)^{2}f(u)}+\frac{d\upsilon
^{2}}{(\upsilon -u)^{2}(\upsilon -w)^{2}\varphi (\upsilon )}+\frac{dw^{2}}{%
(w-u)^{2}(w-\upsilon )^{2}\psi (w)},  \label{m}
\end{equation}%
is no longer flat for generic $f(u)$, $\varphi (\upsilon )$ and $\psi (w)$.
The commuting flows of system (\ref{bb}) generated by the Hamiltonian
densities $N=u+\upsilon -2w$ and $B=2w^{2}-u^{2}-\upsilon ^{2}$ are
\begin{equation}
u_{\tau }=\frac{1}{2}f^{\prime }u_{x}\text{, \ \ \ }\upsilon _{\tau }=\frac{1%
}{2}\varphi ^{\prime }\upsilon _{x}\text{, \ \ \ \ }w_{\tau }=-\frac{1}{4}%
\psi ^{\prime }w_{x}  \label{n}
\end{equation}%
and%
\begin{equation}
u_{\zeta }=-(2f+uf^{\prime })u_{x}\text{, \ \ \ }\upsilon _{\zeta
}=-(2\varphi +\upsilon \varphi ^{\prime })\upsilon _{x}\text{, \ \ \ \ }%
w_{\zeta }=(\psi +\frac{1}{2}w\psi ^{\prime })w_{x},  \label{b}
\end{equation}%
respectively. To incorporate the times $\tau $ and $\zeta $ into the
reciprocal transformation (\ref{10}), we have to calculate the fluxes $P$, $Q
$ and $R$, $S$ of the Hamiltonian densities $B$ and $N$,%
\begin{eqnarray}
d\tilde{x} &=&Bdx+Adt+Pd\tau +Qd\zeta ,  \label{forms} \\
d\tilde{t} &=&Ndx+Mdt+Rd\tau +Sd\zeta ,  \notag
\end{eqnarray}%
that is, the coefficients which make these $1-$forms closed by virtue of (%
\ref{bb}), (\ref{n}) and (\ref{b}). Here%
\begin{eqnarray*}
Q &=&\frac{1}{2}(\nabla B)^{2}=2[u^{2}f(u)+\upsilon ^{2}\varphi (\upsilon
)+w^{2}\psi (w)], \\
R &=&\frac{1}{2}(\nabla N)^{2}=\frac{1}{2}[f(u)+\varphi (\upsilon )+\psi
(w)],
\end{eqnarray*}%
while $P$ and $S$ are of the form%
\begin{eqnarray*}
P &=&-uf-\upsilon \varphi -w\psi +\overset{u}{\int }f(\nu )d\nu +\overset{%
\upsilon }{\int }\varphi (\nu )d\nu +\overset{w}{\int }\psi (\nu )d\nu , \\
S &=&-uf-\upsilon \varphi -w\psi -\overset{u}{\int }f(\nu )d\nu -\overset{%
\upsilon }{\int }\varphi (\nu )d\nu -\overset{w}{\int }\psi (\nu )d\nu ,
\end{eqnarray*}%
respectively. Notice that $P+S=(\nabla B$, $\nabla N)=g^{ij}B_{i}N_{j}$.
Under the extended reciprocal transformation (\ref{forms}), the commuting
flows (\ref{n}) and (\ref{b})  take the form%
\begin{equation}
u_{\tau }=\mu ^{1}u_{\tilde{x}}\text{, \ \ \ }\upsilon _{\tau }=\mu
^{2}\upsilon _{\tilde{x}}\text{, \ \ \ \ }w_{\tau }=\mu ^{3}w_{\tilde{x}}
\label{newn}
\end{equation}%
and%
\begin{equation}
u_{\zeta }=\lambda ^{1}u_{\tilde{x}}\text{, \ \ \ }\upsilon _{\zeta
}=\lambda ^{2}\upsilon _{\tilde{x}}\text{, \ \ \ \ }w_{\zeta }=\lambda
^{3}w_{\tilde{x}},  \label{newb}
\end{equation}%
where the characteristic speeds are%
\begin{eqnarray*}
\mu ^{1} &=&-\frac{1}{2}(u-\upsilon )(u-w)f^{\prime }-(\upsilon +w)R-P, \\
\mu ^{2} &=&-\frac{1}{2}(\upsilon -u)(\upsilon -w)\varphi ^{\prime
}-(u+w)R-P, \\
\mu ^{3} &=&-\frac{1}{2}(w-u)(w-\upsilon )\psi ^{\prime }-(u+\upsilon )R-P
\end{eqnarray*}%
and%
\begin{eqnarray*}
\lambda ^{1} &=&(u-\upsilon )(u-w)(2f+uf^{\prime })-(\upsilon +w)S-Q, \\
\lambda ^{2} &=&(\upsilon -u)(\upsilon -w)(2\varphi +\upsilon \varphi
^{\prime })-(u+w)S-Q, \\
\lambda ^{3} &=&(w-u)(w-\upsilon )(2\psi +w\psi ^{\prime })-(u+\upsilon )S-Q,
\end{eqnarray*}%
respectively. The transformed flows (\ref{newn}) and (\ref{newb}) commute
with (\ref{9}). According to Theorem 2, the transformed metric (\ref{m}) and
the transformed flows (\ref{newn}) and (\ref{newb}) generate the nonlocal
Hamiltonian operator which, in matrix form, is%
\begin{equation*}
J=\left(
\begin{array}{ccc}
\tilde{g}^{11} & 0 & 0 \\
0 & \tilde{g}^{22} & 0 \\
0 & 0 & \tilde{g}^{33}%
\end{array}%
\right) \frac{d}{_{d\tilde{x}}}+
\end{equation*}%
\begin{equation*}
\left(
\begin{array}{ccc}
\tilde{g}^{11}[\frac{f^{\prime }(u)}{2f(u)}u_{\tilde{x}}+\frac{u_{\tilde{x}%
}-\upsilon _{\tilde{x}}}{u-\upsilon }+\frac{u_{\tilde{x}}-w_{\tilde{x}}}{u-w}%
] & \frac{\tilde{g}^{22}u_{\tilde{x}}+\tilde{g}^{11}\upsilon _{\tilde{x}}}{%
u-\upsilon } & \frac{\tilde{g}^{33}u_{\tilde{x}}+\tilde{g}^{11}w_{\tilde{x}}%
}{u-w} \\
-\frac{\tilde{g}^{22}u_{\tilde{x}}+\tilde{g}^{11}\upsilon _{\tilde{x}}}{%
u-\upsilon } & \tilde{g}^{22}[\frac{\varphi ^{\prime }(\upsilon )}{2\varphi
(\upsilon )}\upsilon _{\tilde{x}}+\frac{u_{\tilde{x}}-\upsilon _{\tilde{x}}}{%
u-\upsilon }+\frac{\upsilon _{\tilde{x}}-w_{\tilde{x}}}{\upsilon -w}] &
\frac{\tilde{g}^{33}\upsilon _{\tilde{x}}+\tilde{g}^{22}w_{\tilde{x}}}{%
\upsilon -w} \\
-\frac{\tilde{g}^{33}u_{\tilde{x}}+\tilde{g}^{11}w_{\tilde{x}}}{u-w} & -%
\frac{\tilde{g}^{33}\upsilon _{\tilde{x}}+\tilde{g}^{22}w_{\tilde{x}}}{%
\upsilon -w} & \tilde{g}^{33}[\frac{\psi ^{\prime }(w)}{2\psi (w)}w_{\tilde{x%
}}+\frac{w_{\tilde{x}}-\upsilon _{\tilde{x}}}{w-\upsilon }+\frac{u_{\tilde{x}%
}-w_{\tilde{x}}}{u-w}]%
\end{array}%
\right) +
\end{equation*}%
\begin{equation*}
\left(
\begin{array}{cccc}
\lambda ^{1}u_{\tilde{x}} & \mu ^{1}u_{\tilde{x}} & \eta ^{1}\upsilon _{%
\tilde{x}} & u_{\tilde{x}} \\
\lambda ^{2}\upsilon _{\tilde{x}} & \mu ^{2}\upsilon _{\tilde{x}} & \eta
^{2}\upsilon _{\tilde{x}} & \upsilon _{\tilde{x}} \\
\lambda ^{3}w_{\tilde{x}} & \mu ^{3}w_{\tilde{x}} & \eta ^{3}w_{\tilde{x}} &
w_{\tilde{x}}%
\end{array}%
\right) (\frac{d}{d\tilde{x}})^{-1}\left(
\begin{array}{ccc}
u_{\tilde{x}} & \upsilon _{\tilde{x}} & w_{\tilde{x}} \\
\eta ^{1}u_{\tilde{x}} & \eta ^{2}\upsilon _{\tilde{x}} & \eta ^{3}w_{\tilde{%
x}} \\
\mu ^{1}u_{\tilde{x}} & \mu ^{2}\upsilon _{\tilde{x}} & \mu ^{3}w_{\tilde{x}}
\\
\lambda ^{1}u_{\tilde{x}} & \lambda ^{2}\upsilon _{\tilde{x}} & \lambda
^{3}w_{\tilde{x}}%
\end{array}%
\right) ,
\end{equation*}%
where $\tilde{g}^{ii}$ are components of the transformed metric (\ref{m}), $%
\eta ^{1}=\upsilon +w$, $\eta ^{2}=u+w$ and $\eta ^{3}=u+\upsilon $ are the
characteristic speeds of system (\ref{9}), and $\lambda ^{k}$, $\mu ^{k}$
are the characteristic speeds of its commuting flows (\ref{newn}) and (\ref%
{newb}). In the particular case $f(u)=\alpha $, $\varphi (\upsilon )=\beta $
and $\psi (w)=\gamma $, where $\alpha $, $\beta $, $\gamma $ are constants
subject to a single constraint $\alpha +\beta +\gamma =0$, we have $\lambda
^{k}=\mu ^{k}=0$, so that the transformed operator takes the local
Dubrovin-Novikov form, the corresponding flat metric being%
\begin{equation*}
d\tilde{s}^{2}=\frac{du^{2}}{\alpha (u-\upsilon )^{2}(u-w)^{2}}+\frac{%
d\upsilon ^{2}}{\beta (\upsilon -u)^{2}(\upsilon -w)^{2}}+\frac{dw^{2}}{%
\gamma (w-u)^{2}(w-\upsilon )^{2}}.
\end{equation*}%
As shown in \cite{max87}, these are the only local Hamiltonian structures of
hydrodynamic type of the $3-$component system (\ref{9}).
 
\textbf{Remark}. Modifying the reciprocal transformation (\ref{10}) from
Example 3 as%
\begin{eqnarray*}
d\tilde{x} &=&[2R(w)-P(u)-Q(\upsilon )]dx+[Q(\upsilon )-P(u)]dt, \\
d\tilde{t} &=&[A(u)+B(\upsilon )-2C(w)]dx+[A(u)-B(\upsilon )]dt,
\end{eqnarray*}%
we obtain the transformed system%
\begin{equation}
u_{\tilde{t}}=\frac{R(w)-Q(\upsilon )}{C(w)-B(\upsilon )}u_{\tilde{x}}\text{%
, \ \ \ }\upsilon _{\tilde{t}}=\frac{P(u)-R(w)}{A(u)-C(w)}\upsilon _{\tilde{x%
}}\text{, \ \ \ }w_{\tilde{t}}=\frac{Q(\upsilon )-P(u)}{B(\upsilon )-A(u)}w_{%
\tilde{x}},  \label{gn}
\end{equation}%
where $A(u)$, $B(\upsilon )$, $C(w)$ and $P(u)$, $Q(\upsilon )$, $R(w)$ are
arbitrary functions. As shown in \cite{fr91a}, formulae (\ref{gn}) define a
general $3-$component linearly degenerate semi-Hamiltonian system. Repeating
the construction of Example 3, one can obtain the associated family of
nonlocal Hamiltonian structures.
 
\section{Generalized reciprocal transformations}
 
Consider $N$ commuting flows of hydrodynamic type%
\begin{equation}
R_{t^{\beta }}^{i}=\upsilon _{\beta }^{i}(R)R_{t^{1}}^{i}\text{, \ \ \ }%
i=1,2,..., n\text{, \ \ \ }\beta =1,2,..., N,  \label{set}
\end{equation}%
where we have set $t^{1}\equiv x$ and $t^{2}\equiv t$ (so that $\upsilon
_{1}^{i}\equiv 1$). We assume that the flows (\ref{set}) are Hamiltonian
with the local Hamiltonian operator (\ref{ham}) generated by the flat metric
$ds^{2}=g_{ii}(R)dR^{i^{2}}$, so that $\upsilon _{\beta }^{i}=\nabla
^{i}\nabla _{i}h_{\beta }$, where $h_{\beta }$ are the corresponding
Hamiltonian densities. Let us change from $t^{1}$, $t^{2}$,..., $t^{N}$ to
the new independent variables $\tilde{t}^{1}$, $\tilde{t}^{2}$,..., $\tilde{t%
}^{N}$ defined as%
\begin{equation}
d\tilde{t}^{\gamma }=a_{\beta }^{\gamma }(R)dt^{\beta }\text{, \ \ \ }\gamma
,\beta =1,2,...N,  \label{huge}
\end{equation}%
where $a_{1}^{\gamma }(R)$ are the conserved densities of systems (\ref{set}%
), and $a_{\beta }^{\gamma }(R)$ are the corresponding fluxes:%
\begin{equation}
\partial _{t^{\beta }}a_{1}^{\gamma }=\partial _{t^{1}}a_{\beta }^{\gamma }.
\label{bigo}
\end{equation}%
Transformations of the type (\ref{huge}) naturally generalize reciprocal
transformations (the case $N=2$). Under the \textit{generalized} reciprocal
transformation (\ref{huge}), the commuting flows (\ref{set}) transforms to%
\begin{equation}
R_{\tilde{t}^{\beta }}^{i}=\tilde{\upsilon}_{\beta }^{i}(R)R_{\tilde{t}%
^{1}}^{i},  \label{setmod}
\end{equation}%
where%
\begin{equation}
\tilde{\upsilon}_{\beta }^{i}=\frac{A_{\beta }^{\gamma }\upsilon _{\gamma
}^{i}}{A_{1}^{\nu }\upsilon _{\nu }^{i}},  \label{vela}
\end{equation}%
while the metric $ds^{2}$ transforms to
\begin{equation}
d\tilde{s}^{2}=\tilde{g}_{ii}dR^{i^{2}}{}\text{, \ \ \ \ \ \ }\tilde{g}%
_{ii}=g_{ii}(A_{1}^{\nu }\upsilon _{\nu }^{i})^{2}.  \label{meta}
\end{equation}%
Here $A_{\gamma }^{\beta }$ is the inverse of $a_{\beta }^{\nu }$,%
\begin{equation}
dt^{\beta }=A_{\gamma }^{\beta }d\tilde{t}^{\gamma }.  \label{invrec}
\end{equation}%
We emphasize that the transformed metric $d\tilde{s}^{2}$ is no longer flat
in general. In the particular case $N=2$ formula (\ref{meta}) reduces to (%
\ref{new}). To calculate the curvature tensor of the metric $d\tilde{s}^{2}$
we introduce $N$ extra flows%
\begin{equation}
R_{\tau ^{\gamma }}^{i}=w_{\gamma }^{i}(R)R_{t^{1}}^{i}\text{, \ \ \ }%
i=1,2,..., n\text{, \ \ \ }\gamma =1,2,..., N,  \label{seta}
\end{equation}%
generated by the Hamiltonian operator (\ref{ham}) and the Hamiltonian
densities $a_{1}^{\beta }(R)$:%
\begin{equation*}
w_{\beta }^{i}=\nabla ^{i}\nabla _{i}a_{1}^{\beta }.
\end{equation*}%
Clearly, the flows (\ref{seta}) commute with (\ref{set}). To calculate the
transformed flows (\ref{seta}), we have to incorporate the times $\tau
^{\beta }$ into the generalized reciprocal transformation (\ref{huge}),
namely,%
\begin{equation*}
d\tilde{t}^{\gamma }=a_{\beta }^{\gamma }dt^{\beta }+c_{\beta }^{\gamma
}d\tau ^{\beta }\text{, \ \ \ }\gamma ,\beta =1,2,..., N.
\end{equation*}%
Here the fluxes $c_{\beta }^{\gamma }$ are restricted by%
\begin{equation}
c_{\gamma }^{\beta }+c_{\beta }^{\gamma }=(\nabla a_{1}^{\beta }\text{, }%
\nabla a_{1}^{\gamma })=\sum g^{kk}(\partial _{k}a_{1}^{\beta })(\partial
_{k}a_{1}^{\gamma }),  \label{compat}
\end{equation}%
where $(\nabla f$, $\nabla g)$ denotes a scalar product of the gradients of $f$
and $g$ in the metric $ds^{2}$. Finally, the transformed flows (\ref{seta}%
) take the form%
\begin{equation}
R_{\tau ^{\gamma }}^{i}=\tilde{w}_{\gamma }^{i}(R)R_{\tilde{t}^{1}}^{i},
\label{setamod}
\end{equation}%
with%
\begin{equation}
\tilde{w}_{\gamma }^{i}=\frac{w_{\gamma }^{i}-c_{\gamma }^{\beta }A_{\beta
}^{\varepsilon }\upsilon _{\varepsilon }^{i}}{A_{1}^{\nu }\upsilon _{\nu
}^{i}}.  \label{wela}
\end{equation}
 
\begin{theorem}
The curvature tensor of the transformed metric $d\tilde{s}^{2}$ is%
\begin{equation*}
\tilde{R}_{ij}^{ij}=\overset{N}{\underset{\beta =1}{\sum }}\left( \tilde{%
\upsilon}_{\beta }^{i}\tilde{w}_{\beta }^{j}+\tilde{\upsilon}_{\beta }^{j}%
\tilde{w}_{\beta }^{i}\right) .
\end{equation*}%
The metric $d\tilde{s}^{2}$ generates the nonlocal Hamiltonian operator%
\begin{equation*}
\tilde{J}^{ij}=\tilde{g}^{ii}\delta ^{ij}\frac{d}{d\tilde{x}}-\tilde{g}^{ii}%
\tilde{\Gamma}_{ik}^{j}R_{\tilde{x}}^{k}+\overset{N}{\underset{\beta =1}{%
\sum }}\tilde{\upsilon}_{\beta }^{i}R_{\tilde{x}}^{i}\left( \frac{d}{d\tilde{%
x}}\right) ^{-1}\tilde{w}_{\beta }^{j}R_{\tilde{x}}^{j}+\overset{N}{\underset%
{\beta =1}{\sum }}\tilde{w}_{\beta }^{i}R_{\tilde{x}}^{i}\left( \frac{d}{d%
\tilde{x}}\right) ^{-1}\tilde{\upsilon}_{\beta }^{j}R_{\tilde{x}}^{j}.
\end{equation*}
\end{theorem}
 
\ \ \ \ \ \ \ \ \ \ \ \ \ \ \ \ \ \ \ \ \ \ \ \ \ \ \ \ \ \ \ \ \ \ \ \ \ \
\ \ \ \ \ \ \ \ \ \ \ \ \ \textbf{Proof}:
 
let us introduce the Lame coefficients $H_{1i}=\sqrt{g_{ii}}$ and the
rotation coefficients $\beta _{ik}=\partial _{i}H_{1k}/H_{1i}$ (for any $%
i\neq k$). According to (\cite{ts91}), the linear problem%
\begin{equation*}
\partial _{i}H_{k}=\beta _{ik}H_{i}, ~~~~ i\neq k,
\end{equation*}%
has $N$ particlular solutions $H_{\beta i}$, where $H_{\beta i}=\upsilon
_{\beta }^{i}H_{1i}$, and $N$ other solutions $H_{i}^{\beta }$, where $%
H_{i}^{\beta }=w_{\beta }^{i}H_{1i}$. Since commuting flows (\ref{set}) and (%
\ref{seta}) are locally Hamiltonian ($\upsilon _{\beta }^{i}=\nabla
^{i}\nabla _{i}h_{\beta }$ and $w_{\beta }^{i}=\nabla ^{i}\nabla
_{i}a_{1}^{\beta }$), the relationship between conserved densities ($%
h_{\beta }$ and $a_{1}^{\beta }$) and the corresponding commuting flows ($%
\upsilon _{\beta }^{i}$ and $w_{\beta }^{i}$) can be written as%
\begin{equation}
H_{\beta  i}=\psi _{\beta i,i}+\underset{m\neq i}{\sum }\beta _{mi}\psi
_{\beta m}\text{, \ \ \ \ \ }H_{i}^{\beta }=\psi _{i,i}^{\beta }+\underset{%
m\neq i}{\sum }\beta _{mi}\psi _{m}^{\beta },  \label{hamm}
\end{equation}%
where $\psi _{\beta i}$ and $\psi _{i}^{\beta }$ are defined by the
equations $\partial _{i}h_{\beta }=\psi _{\beta i}H_{1i}$ and $\partial
_{i}a_{1}^{\beta }=\psi _{i}^{\beta }H_{1i}$. In this notation the formulas (%
\ref{vela}), (\ref{wela}) and (\ref{meta}) become%
\begin{equation*}
\tilde{H}_{\beta i}=A_{\beta }^{\gamma }H_{\gamma i}\text{, \ \ \ }\tilde{H}%
_{i}^{\beta }=H_{i}^{\beta }-\tilde{H}_{\gamma i}c_{\beta }^{\gamma }
\end{equation*}%
(where $\tilde{\upsilon}_{\beta }^{i}=\tilde{H}_{\beta i}/\tilde{H}_{1i}$
and $\tilde{w}_{\beta }^{i}=\tilde{H}_{i}^{\beta }/\tilde{H}_{1i}$) and the
transformed rotation coefficients are%
\begin{equation*}
\tilde{\beta}_{ik}=\beta _{ik}-\psi _{i}^{\beta }\tilde{H}_{\beta  k}.
\end{equation*}%
Since the metric $ds^{2}$ is flat, the curvature components $R_{jik}^{i}$ ($%
i\neq j\neq k$) vanish identically, that is, $\partial _{i}\beta _{jk}=\beta
_{ji}\beta _{ik}$ ($i\neq j\neq k$). This identity is preserved under
generalized reciprocal transformations: $\partial _{i}\tilde{\beta}_{jk}=%
\tilde{\beta}_{ji}\tilde{\beta}_{ik}$. However, the components $\tilde{R}%
_{ij}^{ij}$ defined as%
\begin{equation*}
\tilde{R}_{ij}^{ij}=-\frac{\tilde{\Delta}_{ij}}{\tilde{H}_{1i}\tilde{H}_{1j}}
\end{equation*}%
where%
\begin{equation*}
\tilde{\Delta}_{ij}\equiv \partial _{i}\tilde{\beta}_{ij}+\partial _{j}%
\tilde{\beta}_{ji}+\underset{m\neq i}{\sum }\tilde{\beta}_{mi}\tilde{\beta}%
_{mj},
\end{equation*}%
will no longer be zero. Indeed,
\begin{equation*}
\begin{array}{c}
\tilde{\Delta}_{ij}=\partial _{i}\tilde{\beta}_{ij}+\partial _{j}\tilde{\beta%
}_{ji}+\underset{m\neq i\neq k}{\sum }\tilde{\beta}_{mi}\tilde{\beta}_{mj}=
\\
=\partial _{i}[\beta _{ij}-\psi _{i}^{\gamma }\tilde{H}_{\gamma j}]+\partial
_{j}[\beta _{ji}-\psi _{j}^{\gamma }\tilde{H}_{\gamma i}]+\underset{m\neq
i\neq j}{\sum }(\beta _{mi}-\psi _{m}^{\gamma }\tilde{H}_{\gamma i})(\beta
_{mj}-\psi _{m}^{\beta }\tilde{H}_{\beta j}) \\
=\Delta _{ij}-\tilde{H}_{\gamma j}[\psi _{i,i}^{\gamma }+\underset{m\neq i}{%
\sum }\beta _{mi}\psi _{m}^{\gamma }]-\tilde{H}_{\gamma i}[\psi
_{j,j}^{\gamma }+\underset{m\neq j}{\sum }\beta _{mj}\psi _{m}^{\gamma }]+%
\tilde{H}_{\beta i}\tilde{H}_{\gamma j}\sum \psi _{m}^{\beta }\psi
_{m}^{\gamma }.%
\end{array}%
\end{equation*}%
Since $\Delta _{ij}\equiv 0$, equations (\ref{hamm}) and (\ref{compat}),
imply%
\begin{eqnarray*}
\tilde{\Delta}_{ij} &=&-\tilde{H}_{\gamma j}H_{i}^{\gamma }-\tilde{H}%
_{\gamma i}H_{j}^{\gamma }+\tilde{H}_{\gamma i}\tilde{H}_{\beta j}(c_{\gamma
}^{\beta }+c_{\beta }^{\gamma })= \\
&=&-\tilde{H}_{\beta j}[H_{i}^{\beta }-\tilde{H}_{\gamma i}c_{\beta
}^{\gamma }]-\tilde{H}_{\beta j}[H_{i}^{\beta }-\tilde{H}_{\gamma i}c_{\beta
}^{\gamma }]= \\
&=&-\tilde{H}_{\beta j}\tilde{H}_{i}^{\beta }-\tilde{H}_{\beta i}\tilde{H}%
_{j}^{\beta }.
\end{eqnarray*}%
Thus, the transformed curvature tensor is%
\begin{equation*}
\tilde{R}_{ij}^{ij}=-\frac{1}{\tilde{H}_{1i}\tilde{H}_{1j}}\tilde{\Delta}%
_{ij}=\frac{\tilde{H}_{\beta j}\tilde{H}_{i}^{\beta }+\tilde{H}_{\beta i}%
\tilde{H}_{j}^{\beta }}{\tilde{H}_{1i}\tilde{H}_{1j}}=\tilde{\upsilon}%
_{\beta }^{i}\tilde{w}_{\beta }^{j}+\tilde{\upsilon}_{\beta }^{j}\tilde{w}%
_{\beta }^{i}.
\end{equation*}%
For $N=2$ this proves Theorem 2 formulated in section 4.
 
\bigskip

As shown in \cite{fr91a}, any linearly degenerate semi-Hamiltonian system in 
Riemann invariants can be linearized by a generalized reciprocal transformation.
Applying the results of section 5 to local Hamiltonian structures of a 
linear system, one can explicitly construct an infinity of mutually compatible 
nonlocal Hamiltonian structures for an arbitrary linearly degenerate 
semi-Hamiltonian system in the same way as it was done in examples 1-3.
In the example below we explicitly construct nonlocal Hamiltonian formalism for the n-component linearly degenerate system mentioned in the abstract.

\textbf{Example 4}.

As shown in \cite{fr91a}, the linearly degenerate semi-Hamiltonian system%
\begin{equation}
R_{\tilde{t}^{2}}^{i}=\left( R^{i}-\sum_{m=1}^n R^{m}\right) R_{\tilde{t}^{1}}^{i}%
\text{, \ \ \ }i=1,2,..., n,  \label{nxn}
\end{equation}%
and its linearly degenerate commuting flows%
\begin{equation*}
R_{\tilde{t}^{\beta }}^{i}=(-1)^{\beta +1}\left( \partial _{i}\tilde{h}_{\beta }\right) R_{%
\tilde{t}^{1}}^{i}\text{, \ \ \ }i=1,2,..., n\text{, \ \ \ \ }\beta =3\text{, }%
4\text{, }..., n,
\end{equation*}%
where%
\begin{equation*}
\tilde{h}=\underset{k=1}{\overset{n}{\prod }}(1+\lambda R^{k})=1+\lambda 
\tilde{h}_{1}+\lambda ^{2}\tilde{h}_{2}+\lambda ^{3}\tilde{h}%
_{3}+...+\lambda ^{n}\tilde{h}_{n},
\end{equation*}%
can be obtained from a set of commuting linear flows %
\begin{equation*}
R_{t^{\beta }}^{i}=(\varepsilon ^{i})^{\beta -1}R_{t^{1}}^{i}\text{, \ \ \ \
\ \ }\beta =1\text{, }2\text{, }..., n,
\end{equation*}%
($\varepsilon ^{i}$ are
arbitrary constants) by a generalized reciprocal transformation, %
\begin{equation*}
d\tilde{t}^{\beta }=a_{\gamma }^{\beta }dt^{\gamma },
\end{equation*}%
where%
\begin{equation*}
a_{\gamma }^{\beta }=\underset{k=1}{\overset{n}{\sum }}(\varepsilon
^{k})^{\gamma -1}(R^{k})^{n-\beta }.
\end{equation*}%
In the $2-$ and $3-$component cases
the explicit form of this transformation is stated in Examples 1 and 3. The inverse reciprocal transformation is%
\begin{equation*}
dt^{\beta }=A_{\gamma }^{\beta }d\tilde{t}^{\gamma },
\end{equation*}%
where%
\begin{equation*}
A_{\gamma }^{\beta }=\underset{k=1}{\overset{n}{\sum }}(-1)^{n+\gamma +\beta
+1}\frac{\partial _{k}\tilde{\varepsilon}_{n+1-\gamma }}{\underset{m\neq k}{%
\overset{n}{\prod }}(\varepsilon ^{k}-\varepsilon ^{m})}\frac{\partial _{k}%
\tilde{h}_{\beta }}{\underset{s\neq k}{\overset{n}{\prod }}(R^{k}-R^{s})}
\end{equation*}%
and%
\begin{equation*}
\tilde{\varepsilon}=\underset{k=1}{\overset{n}{\prod }}(1+\lambda
\varepsilon ^{k})=1+\lambda \tilde{\varepsilon}_{1}+\lambda ^{2}\tilde{%
\varepsilon}_{2}+\lambda ^{3}\tilde{\varepsilon}_{3}+...+\lambda ^{n}\tilde{%
\varepsilon}_{n}.
\end{equation*}%
Applying Theorem 4 to local Hamiltonian structures of the corresponding
linear systems, we obtain infinitely many nonlocal Hamiltonian operators
associated with the system (\ref{nxn}):%
\begin{equation}
A^{ij}=g^{ii}\delta ^{ij}\frac{d}{dx}-g^{ii}\Gamma _{ik}^{j}R_{x}^{k}+%
\overset{n-1}{\underset{\beta =1}{\sum }}\left[ \tilde{\upsilon}_{\beta
}^{i}R_{x}^{i}\left( \frac{d}{dx}\right) ^{-1}\tilde{w}_{\beta
}^{j}R_{x}^{j}+\tilde{w}_{\beta }^{i}R_{x}^{i}\left( \frac{d}{dx}\right)
^{-1}\tilde{\upsilon}_{\beta }^{j}R_{x}^{j}\right].   \label{i}
\end{equation}%
Here $g_{ii}(dR^i)^2$ is the diagonal metric of the form%
\begin{equation}
\frac{dR^{1^{2}}}{\underset{k\neq 1}{\prod }(R^{1}-R^{k})f^{1}(R^{1})}+...+%
\frac{dR^{n^{2}}}{\underset{k\neq n}{\prod }(R^{n}-R^{k})f^{n}(R^{n})},
\label{nmetr}
\end{equation}%
with the curvature tensor
\begin{equation*}
\tilde{R}_{ik}^{ik}=\overset{n-1}{\underset{\beta =1}{\sum }}\left( \tilde{%
\upsilon}_{\beta }^{i}\tilde{w}_{\beta }^{k}+\tilde{\upsilon}_{\beta }^{k}%
\tilde{w}_{\beta }^{i}\right) ,
\end{equation*}%
where%
\begin{eqnarray*}
\tilde{\upsilon}_{\beta }^{i} &=&\partial _{i}\tilde{h}_{\beta }\text{,} \\
\tilde{w}_{\beta }^{i} &=&(n-\beta )\underset{k\neq i}{\prod }(R^{i}-R^{k})%
\left[ \frac{1}{2}f_{i}^{\prime }(R^{i})(R^{i})^{n-\beta -1}+(n-\beta
-1)f_{i}(R^{i})(R^{i})^{n-\beta -2}\right] -\overset{n-1}{\underset{\gamma =1%
}{\sum }}\tilde{\upsilon}_{\gamma }^{i}c_{\beta }^{\gamma }, \\
\partial _{i}c_{\beta }^{\gamma } &=&(n-\beta )(n-\gamma )\left[ \frac{1}{2}%
f_{i}^{\prime }(R^{i})(R^{i})^{2n-\beta -\gamma -2}+(n-\beta
-1)f_{i}(R^{i})(R^{i})^{2n-\beta -\gamma -3}\right] .
\end{eqnarray*}%
Operators (\ref{i}) depend on $n$ arbitrary functions $f^{1}(R^{1})$, ..., $%
f^{n}(R^{n})$, and are mutually compatible. It was demonstrated in \cite%
{max87} that the metric (\ref{nmetr}) is not flat for $n\geq 4$, whatever $%
f^{i}(R^{i})$ are.

\section{Acknowledgements}
 
We thank the Royal Society for the their financial support of MVP to
Loughborough, making this collaboration possible. MVP is also partially supported by the
Russian Foundation for Fundamental
Research (Grants 00-01-00210 and 00-01-00366).

\end{document}